\documentclass{ametsocV6.1_no_lineno}

\usepackage{comment}
\usepackage{enumitem}
\usepackage{siunitx}
\usepackage{stfloats}   

\title{Southern Ocean latent heat flux variability \\driven by oceanic meso- and submesoscale motions}

\authors{Lucie Reymondet,\aff{a}\correspondingauthor{Lucie Reymondet, lureymondet@ucsd.edu}
Lia Siegelman,\aff{a}
and Luc Lenain,\aff{a}
}

\affiliation{\aff{a}{Scripps Institution of Oceanography, UC San Diego, La Jolla, CA, USA}}

\abstract{Latent heat flux is a primary pathway for ocean–atmosphere exchange of heat and moisture, yet the influence of sea surface temperature variability at fine scales ($\leq$ 100 km) on latent heat flux variability, particularly over the Southern Ocean, remains poorly understood.
Here we quantify the scale-dependent drivers of latent heat flux (LHF) variability using a year-long, global, fully coupled ocean–atmosphere simulation with kilometer-scale resolution.
Annual-mean LHF in eddy-rich regions reaches $\approx$ 215 W m$^{-2}$, approximately three times larger than in eddy-poor regions.
Spectral analyses show that ocean mesoscale [$\mathcal{O}$(100 km)] and submesoscale [$\mathcal{O}$(1–10 km)] variability accounts for up to $\approx$ 80\% of the total LHF variance in eddy-rich sectors, but as little as 10\% in eddy-poor regions, and increases proportionally with eddy kinetic energy and sea surface temperature (SST) variance.
We also find that strong submesoscale SST fronts ($\approx$ 5 $^\circ$C over 10 km) force a localized secondary circulation that extends well above the marine boundary layer into the mid-troposphere.
Comparison with ERA5 shows that fine ocean scales, responsible for about 17\% of the ocean-driven LHF variance in the simulation, are largely unresolved in the reanalysis, leading to a muted atmospheric response lacking any secondary circulation. 
Despite a strong heterogeneity in LHF variability, the atmospheric dynamics are mostly uniform across the domain, suggesting a non local atmospheric response to ocean forcing.
These results highlight the potential for ocean meso- and submesoscales, commonly under-resolved in climate models and reanalysis, to influence Southern Ocean air–sea coupling and atmosphere both locally and remotely. 
}

\begin{document}

\maketitle

%
%
\statement
The Southern Ocean plays a key role in regulating Earth’s climate by exchanging heat and moisture with the atmosphere. Using a groundbreaking, high-resolution global simulation of the ocean and atmosphere down to km-scale, we show that small ocean features - eddies and sharp temperature fronts tens to just a few kilometers wide - dominate latent heat flux variability in energetic regions. These fine ocean structures can trigger strong, narrow updrafts that reach up to the mid-troposphere. This small-scale atmospheric response is largely missing from widely used, coarser reanalyses and climate models. Because these fine scale ocean processes can influence the atmosphere locally and remotely, properly representing them is essential for improving climate projection and weather forecast.

%


\section{Introduction}
The ocean is a major heat reservoir in the Earth system, and growing evidence has shown that air-sea exchanges at mesoscale can substantially influence large-scale oceanic and atmospheric circulation \citep{seo2023ocean}. Latent heat fluxes (LHF) provide a primary pathway for the exchange of heat and moisture between the ocean and the atmosphere, as they capture the bulk of turbulent heat fluxes, far exceeding the amplitude of sensible heat fluxes over warm sea surface temperature (SST) anomalies \citep{small2019air,tamsitt2020mooring}.

In fact, ocean mesoscales ($\sim$100-500 km scale) account for more than half of the monthly LHF variance in western boundary currents (WBCs) and in the Antarctic Circumpolar Current (ACC) \citep{bishop2017scale,small2019air}. 
To understand the impact of ocean mesoscales on LHF, it is useful to consider the physical mechanisms at play \cite[see for example Fig.~1 in][]{seo2023ocean}. When an air mass flows across a SST gradient, it creates air–sea differences in temperature and humidity. Wind blowing from cool to warm SST produces an imbalance that generates a positive turbulent heat flux anomaly on the warm side of the front, enhancing ocean heat loss\footnote{Note that throughout this paper, we use the convention that turbulent heat fluxes are positive out of the ocean, i.e., ocean cooling.}. This thermal imbalance deepens and destabilizes the boundary layer, enhancing turbulent mixing, decreasing wind shear and increasing surface wind stress. In the warm-to-cool case, the effects are reversed. Steeper SST gradients and stronger winds further accentuate the air-sea thermodynamic disequilibrium. In addition, recent observational and numerical studies show that submesoscale fronts of SST enhance vertical mixing, diabatic processes, and promote cloud formation and precipitations \citep{kaouah2025submesoscale,yang2024observations, vivant2025meandering, vivant2025ocean,strobach2022local}. 
Yet identifying the drivers of LHF variability remains challenging due to the complex interactions between the ocean and the atmosphere occurring over a broad range of spatial and temporal scales (from hours to months and kilometers to thousands of kilometers, Fig.~\ref{fig-key}).

The Southern Ocean (SO) offers an ideal natural laboratory to address this question as it features a wide range of oceanic dynamical regimes, from large-scale currents to mesoscale eddies and submesoscale filaments, to intense wind events. The SO comprises areas of elevated eddy kinetic energy (EKE) \citep{thompson2014equilibration, beech2022long,meijer2022dynamics}, as well as regions of weak EKE \citep{rintoul2018global}. Regions of high EKE feature abundant mesoscale eddies and submesoscale fronts, concentrated along the ACC and near its confluences with the Agulhas and Brazil Currents \citep{rosso2014vertical, siegelman2020energetic,siegelman2020enhanced,taylor2023submesoscale,marshall2012closure}. The SO also experiences some of the strongest near-surface winds on Earth \citep{sampe2007satellitemapping}, which modulate air-sea heat fluxes. 
Despite the region's pivotal role in ocean-atmosphere heat exchanges \citep{morrison2022ventilation, gruber2019variable, frolicher2015dominance, williams2023role}, latent heat fluxes in the SO remain poorly constrained, largely due to the scarcity of observations imposed by harsh remote conditions that restrict in situ measurements and persistent cloud cover that limits optical satellite observations \citep{prend2025observing}.

Previous studies have begun to explore the variability of SO surface heat fluxes. Using the European Centre for Medium-Range Weather Forecasts ERA5 atmospheric reanalysis \citep{hersbach2020era5} and a data-assimilating ocean model, \citet{prend2025observing} established statistics of net air–sea flux variability across the SO. Their study reveals pronounced geographical heterogeneity, yet the horizontal resolutions in the atmosphere (1/4$^\circ$) and ocean (1/6$^\circ$) were too coarse to resolve mesoscale and submesoscale processes at high latitudes, while SO air-sea flux uncertainties in reanalysis products are known to be largest in regions of high mesoscale activity \citep{swart2019constraining}.
At even finer scale, recent \textit{in situ} observations collected by Saildrones in the Pacific sector of the SO show that submesoscale SST fronts of order $\mathcal{O}$(km) can contribute significantly to sensible heat flux variability, with fronts less than 1 km wide accounting for twice the sensible heat flux variability induced by those larger than 4 km \citep{edholm2025synoptic}, underscoring the need to quantify the contribution of ocean mesoscales and submesoscales, as well as atmospheric dynamics, to LHF variability across the Southern Ocean.

Here, we use a groundbreaking global and year-long coupled ocean-atmosphere simulation (COAS) with a resolution of 2-4 km in the ocean and 6-7 km in the atmosphere (see methods). COAS represents a major step forward by making it possible to resolve fine scale processes simultaneously across the ocean–atmosphere interface. Two recent studies using COAS in the Gulf Stream \citep{strobach2022local} and the Kuroshio Extension \citep{vivant2025ocean} have shown a characteristic atmospheric response to submesoscale fronts. Strong SST gradients ($\sim$1-5 $^\circ$C 10 km$^{-1}$) were found to trigger a collocated secondary vertical circulation leading to narrow enhanced diabatic processes and convective precipitations on the horizontal scale of the submesoscale front, suggesting that similar fine scale coupled dynamics may be at play in the Southern Ocean.

The paper is structured as follows. Section~\ref{data-methods} introduces the COAS and ERA5 datasets and the methods of analysis. Section~\ref{heterogeneity-SO} describes the spatial heterogeneity of latent heat fluxes in the Southern Ocean. Section~\ref{disentangling-physical} investigates the drivers of LHF variability in physical space, by examining the distributions of oceanic and atmospheric variables regulating bulk latent heat fluxes. Section~\ref{disentangling-spectral} moves to spectral space, using wavenumber-frequency analysis to partition between ocean- and atmosphere-driven contributions to LHF variance. Section~\ref{sms-fronts} examines the role of submesoscale fronts in modulating latent heat fluxes and presents a case study illustrating the mechanisms through which submesoscale fronts can elicit a localized atmospheric response. Section~\ref{ERA5} applies the same diagnostics to ERA5 and discusses limitations of coarse-resolution models in capturing fine scale coupled processes. Our discussion and conclusions appear in section~\ref{conclusion}.

\section{Data and Methods}\label{data-methods}

\subsection{COAS: coupled ocean-atmosphere simulation}\label{methods-COAS}

The coupled model used in this study is the Goddard Earth Observing System (GEOS) atmospheric model coupled to the Massachusetts Institute of Technology general circulation ocean model (MITgcm). GEOS was configured to run with a nominal horizontal grid spacing of 6.9 km and 72 vertical hybrid sigma pressure levels, and MITgcm with a nominal horizontal resolution of 1/24$^\circ$, i.e., 2 to 4 km, and 90 vertical levels.
The coupled ocean-atmosphere simulation (COAS) was run for 14 months. The model outputs contain three-dimensional hourly fields for oceanic and atmospheric variables and 15-minute surface atmospheric fields. More information on the coupled model can be found in \citet{molod2015development} and \citet{torres2022wind}.

The fine temporal and spatial resolution achieved in COAS allow us to explore the full spectrum of coupled air-sea processes across a broad range of scales, down to kilometer and hourly variability, representing a substantial advance over previous numerical studies, that typically used resolutions of 0.1$^\circ$ to 0.25$^\circ$ in the ocean, and 0.25$^\circ$ in the atmosphere \citep{small2019air,bishop2017scale,chang2020unprecedented,prend2025observing}. For the purpose of collocating oceanic and atmospheric variables, the simulation outputs are interpolated from the respective model grids onto a regular grid with 0.04$^\circ$ spacing in latitude and longitude.

Surface turbulent fluxes are parameterized using a modified version of \citet{helfand1995climatology}, to better capture wind stress at moderate and high wind speeds \citep{garfinkel2011improvement,molod2013impact}. Latent heat flux is expressed using a bulk formulation:
\begin{equation}
\text{LHF} = \rho C_e L_v \left|U_{10}-u\right| \left(q_{SST}-q_{10}\right)
\label{eq:LHF}
\end{equation}
with \(\rho\) the air density, \(C_e\) the turbulent exchange coefficient, \(L_v\) the latent heat of vaporization, \(U_{10}\) and \(q_{10}\) the wind and air humidity 10 meters above the surface, \(u\) and \(q_{SST}\) the ocean surface current and saturated specific humidity (function of SST).

The results shown in this study are based on 11 simulation months from April 1 to February 28. Following \citet{torres2022wind}, the first 2 months of simulation spin-up are discarded. The last month (March) of the simulation is also discarded to retain only continuous seasons (AM, JJA, SON, DJF) for spectral analysis.

\subsection{ERA5 reanalysis}\label{method-era}

The European Centre for Medium-Range Weather Forecasts Reanalysis v5 (ERA5) product \citep{hersbach2020era5} provides hourly estimates of various atmospheric variables. We extract surface variables (SST, 10-m wind, sea-ice cover, hourly mean LHF and wind stress) from 2014 to 2024 in the same domain as the coupled simulation (30$^\circ$S to 80$^\circ$S, 180$^\circ$W to 180$^\circ$E). Note, while the standard ERA5 sign convention for turbulent heat fluxes is positive downward (i.e., ocean warming), we reverse here this convention to align with the one used in COAS, i.e., positive upward. 

For the vertical transect presented in Fig.~\ref{fig-section_ERA}, we extract three-dimensional profiles of specific humidity, temperature, geopotential height, horizontal wind and vertical velocity at pressure levels up to 600 hPa. We then convert vertical velocity with respect to pressure ($\omega$ in Pa s$^{-1}$) to vertical velocity with respect to height ($w$ in m s$^{-1}$) and estimate the density of moist air from specific humidity, pressure, and temperature, using the MetPy library \citep{metpy2022}. The convective mass flux is estimated from the reanalysis fields as $M_c = \rho \times w$, expressed in kg m$^{-2}$ day$^{-1}$, with $\rho$ the air density and $w$ the vertical velocity (Fig.~\ref{fig-section_ERA}e). Note that the resolution-aware convection scheme used in COAS can not be directly applied to ERA5 profiles, as the mismatch in grid resolution ($\sim$30 km in ERA5 against $\sim$7 km in GEOS) would lead to inconsistent representation of the largest convection plumes. The estimate $M_c$ defined above is solely used to illustrate the structure of the convective atmospheric response qualitatively, rather than rigorously quantify its magnitude.

\subsection{Spectral analysis}\label{method-spectral-analysis}

We use wavenumber-frequency $(k-f)$ spectra to quantify the relative contribution of ocean mesoscales and submesoscales, and of atmospheric synoptic scale and high frequencies, to LHF variability. The atmospheric synoptic scale is captured by pressure systems, e.g., storms, with a typical size of 1000 km and a lifespan of 2 to 6 days, while atmospheric high frequencies correspond to transient events, e.g., warm and cold fronts or squall lines, with spatial scales smaller than 200 km and time scales shorter than 1 day \citep{vivant2025ocean}.

Spectral analyses are performed in domains of approximately $\text{1100 km}\times\text{1100 km}$ in size. We use a common isotropic wavenumber grid across all regions (which vary in physical extent and sampling with latitude), with wavenumber spacing $dk = 8\times10^{-4}$ cycles km$^{-1}$ and maximum wavenumber $k_{\max}=0.1$ cycles km$^{-1}$. This choice yields a uniform discretization while retaining the range of scales resolved in both the smallest and largest regions. To map variability across the Southern Ocean, we slide a tile of the same size along a 1$^\circ\times$1$^\circ$ grid, such that tiles centered on adjacent grid points overlap by $\sim$10\%. Any tiles that overlap with land are discarded.

For annual-mean spectra, the 11-month simulation is partitioned into four FFT windows of equal length (83 days), and the resulting spectra are averaged. For seasonal spectra, the window length (and thus the frequency discretization $df$) varies by season because austral autumn (AM) contains one fewer month than austral winter (JJA), spring (SON), and summer (DJF). The seasonal spectra are interpolated onto the same $(k-f)$ grid as the annual-mean spectrum prior to computing seasonal departures.

All frequency-wavenumber spectra and co-spectra are presented in a variance-preserving form, i.e., multiplied by frequency $f$ and wavenumber $k$, to enable visual comparison of the contributions from different time and length scales to the total variance and covariance across the $(k-f)$ domain when plotted in logarithmic scale.

\subsection{Spectral gap identification}\label{method-spectral-gap}

\begin{figure}
\centering
 \noindent\includegraphics{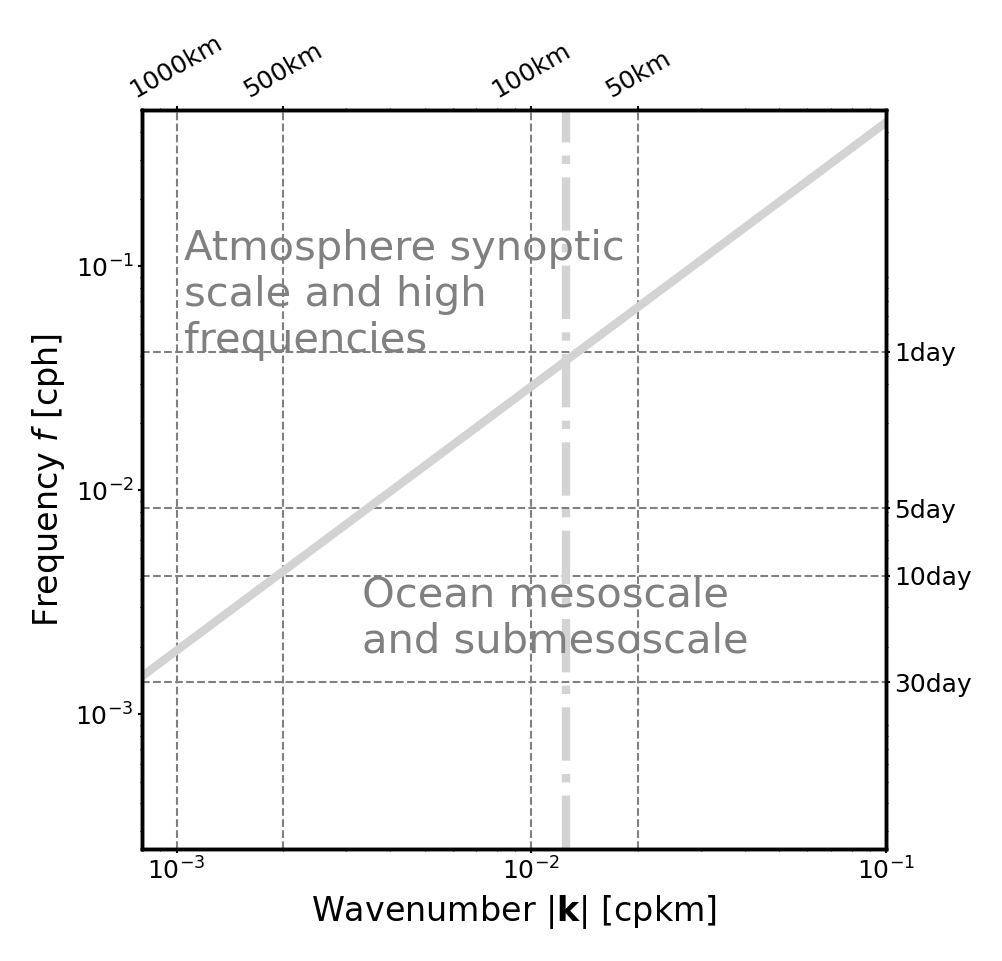}
 \caption{\textbf{Spectral space partition.} Schematic illustrating the spectral space partition that associates each regime with physical scales. The diagonal solid line separates oceanic from atmospheric scales. The dash-dotted vertical line denotes the finest scale ($\sim$80 km) captured by ERA5.}
 \label{fig-key}
\end{figure}

The LHF power spectral density (PSD) shows a distinct diagonal spectral gap in the $(k-f)$ spectral space (Figs.~\ref{fig-key}, \ref{fig-PSDlhf} and appendix Fig.~\ref{fig-PSDlhf-otherregions}). In the Agulhas Retroflection, the Brazil Malvinas Confluence, South of Tasmania and on the Kerguelen Plateau, the LHF spectrum exhibits two prominent maxima, one at high frequencies and low wavenumbers, the other at low frequencies and high wavenumbers (Fig.~\ref{fig-PSDlhf}a and appendix Fig.~\ref{fig-PSDlhf-otherregions}a,b,d). In contrast, in the south Pacific and off the tip of Chile, the LHF spectrum is dominated by high-frequency, low-wavenumber variability, with only a weak contribution at low frequencies and high wavenumbers (Fig.~\ref{fig-PSDlhf}b and appendix Fig.~\ref{fig-PSDlhf-otherregions}c).

This spectral gap separates two regimes: (i) a high-wavenumber, low-frequency regime with spatial scales of $\mathcal{O}$(10–100 km) and time scales of weeks to months (below the diagonal in Fig.~\ref{fig-key}), consistent with the wavenumber–frequency space occupied by oceanic mesoscale and submesoscale variability \citep{torres2018partitioning}; and (ii) a low-wavenumber, high-frequency regime with spatial scales of $\mathcal{O}$(100–1000 km) and time scales shorter than a few days (above the diagonal in Fig.~\ref{fig-key}), consistent with atmospheric synoptic variability and high frequencies.

We exploit this separation to quantify the relative contributions of oceanic and atmospheric variability to latent heat flux variability, building upon \citet{vivant2025ocean}. We integrate the PSD within each regime to estimate its fractional contribution to total variance (see Figs.~\ref{fig-PSDlhf}, \ref{fig-PSDsst}, \ref{fig-seasonlhf}, \ref{fig-PSDlhf_ERA} and appendix Fig.~\ref{fig-PSDlhf-otherregions}). Unlike approaches that diagnose ocean–atmosphere transition scales in space and time separately \citep{bishop2017scale, small2019air, chang2020unprecedented}, wavenumber–frequency spectra capture the coupled variability that spans overlapping spatial and temporal ranges. Details on the partitioning are provided in Appendix B.

\section{Results}

\subsection{Heterogeneity of latent heat fluxes across the Southern Ocean}\label{heterogeneity-SO}

\begin{figure}
 \centering
 \noindent\includegraphics[width = \linewidth]{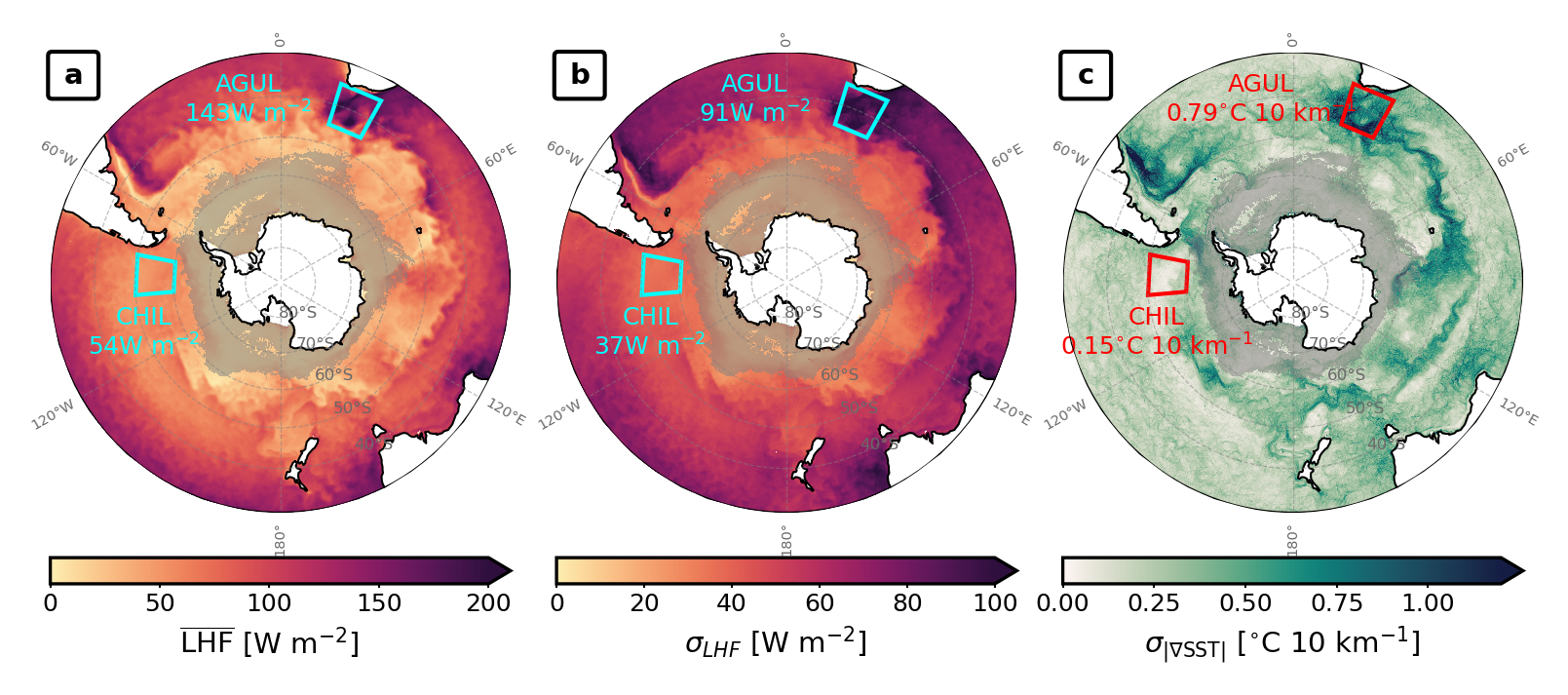}
 \caption{\textbf{Southern Ocean spatial heterogeneity and temporal variability in COAS.} Annual (a) latent heat flux average, (b) latent heat flux standard deviation, and (c) SST gradient standard deviation, all computed over 11 months in COAS. Boxes are AGUL and CHIL regions with regionally averaged values. Gray shaded area is where sea ice coverage exceeds 50\%.}
 \label{fig-SOlhf}
\end{figure}

A large spatial heterogeneity exists on the scale of the basin; annual-mean latent heat fluxes increase with distance from the pole following the large-scale meridional SST gradient (Fig.~\ref{fig-SOlhf}). Annual-mean LHF reach 110 W m$^{-2}$ at 30$^\circ$S, whereas they are only of 22 W m$^{-2}$ at 65$^\circ$S, with an overall mean of 61 W m$^{-2}$. Distinct hotspots appear in both the annual mean and its variability (Fig.~\ref{fig-SOlhf}a,b), notably in the Agulhas Retroflection and Return Current and in the Brazil–Malvinas Confluence. By contrast, fluxes in the southeast Pacific are weaker and less variable, consistent with earlier assessments of net \citep{prend2025observing} and turbulent \citep{liu2011intercomparisons} air–sea heat fluxes.

To further illustrate this heterogeneity, we compare two $\sim$1100 km $\times$ 1100 km regions (boxes in Fig.~\ref{fig-SOlhf}): one in the Agulhas Retroflection (AGUL) and one off the tip of Chile (CHIL). In AGUL, annual-mean latent heat flux locally reaches 214 W m$^{-2}$, whereas it nowhere exceeds 65 W m$^{-2}$ in CHIL. Regional averages show that the mean and standard deviation of LHF are approximately 2.5 times greater in AGUL (143 W m$^{-2}$ and 91 W m$^{-2}$) than in CHIL (54 W m$^{-2}$ and 37 W m$^{-2}$). 

Regions with large LHF variability also exhibit strong SST variability, captured by the variability of the SST gradient (Fig.~\ref{fig-SOlhf}c). In AGUL, the average amplitude of the SST gradient $|\nabla \mathrm{SST}|$ and its standard deviation $\sigma_{|\nabla \mathrm{SST}|}$ are 0.64 $^\circ$C 10 km$^{-1}$ (not shown) and 0.79 $^\circ$C 10 km$^{-1}$, respectively (Fig.~\ref{fig-SOlhf}c). 
In contrast, CHIL is characterized by much weaker gradients: the amplitude of the SST gradient and its standard deviation are 0.14 $^\circ$C 10 km$^{-1}$ (not shown) and 0.15 $^\circ$C 10 km$^{-1}$ (Fig.~\ref{fig-SOlhf}c) on average. The annual standard deviation of the SST gradient amplitude $\sigma_{|\nabla \mathrm{SST}|}$ locally reaches up to 2.2 $^\circ$C 10 km$^{-1}$ in AGUL, comparable in magnitude to submesoscale fronts reported in the Gulf Stream in COAS ($\pm$ 4 $^\circ$C 10 km$^{-1}$) \citep{strobach2022local}.

\begin{figure}
 \centering
 \noindent\includegraphics[width=.85\linewidth]{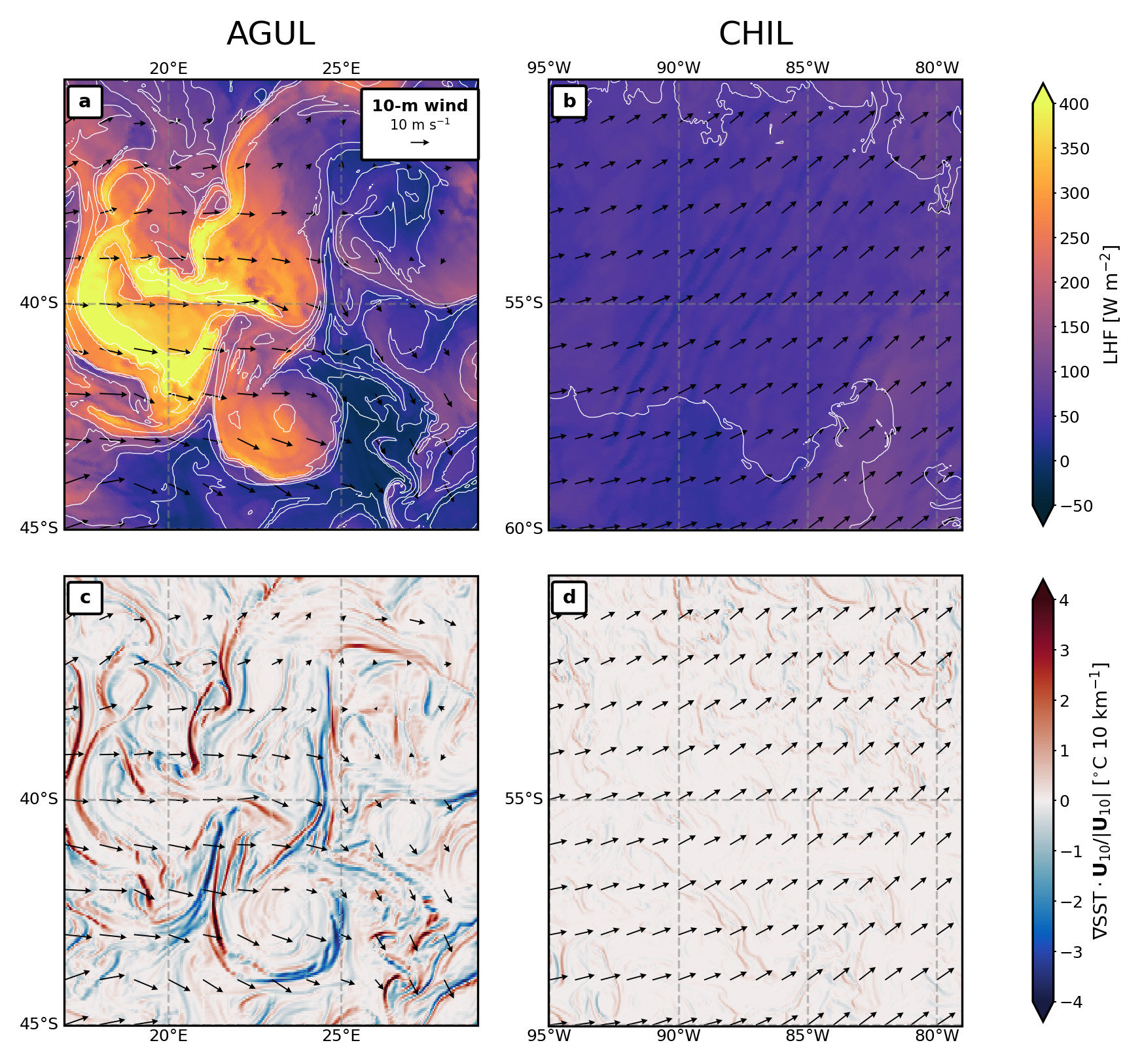}
 \caption{\textbf{Regional snapshots of surface fields in COAS.} (a,b) Latent heat flux (shading) and SST (white contours, spaced 2 $^\circ$C). (c,d) Downwind SST gradient (shading). Arrows represent 10-m horizontal wind. Left column is the AGUL region, right is CHIL. Snapshot from COAS on 1 April 2020, 2300.}
 \label{fig-snapshot}
\end{figure}

The spatial structure of LHF variability is further illustrated in snapshots of the simulation (Fig.~\ref{fig-snapshot}). In the AGUL region, sea surface temperature exhibits a complex mesoscale eddy field with scales of order $\mathcal{O}$(100 km) bordered by sharp submesoscale fronts $\mathcal{O}$(km) wide. Latent heat flux closely tracks the SST field with a pronounced spatial patchiness and large amplitude variations. 
In contrast, in the southeast Pacific, SST gradients are much weaker and smoother, as indicated by widely spaced contours, and the latent heat flux is correspondingly weaker and more spatially uniform. 
In both regions, however, changes in the 10-m wind speed and direction occur primarily on synoptic scale, highlighting that, despite similar wind structure, the LHF response differs markedly between the two snapshots.
These results indicate that LHF variability is strongly coupled to SST variability induced by the ocean, suggesting that fine ocean scales play an important role in setting the heterogeneity of LHF across the SO.

Finally, the temporal scales of LHF variability are best appreciated by examining hourly animations from the simulation (see supplemental material). Latent heat fluxes and their gradient are modulated by atmospheric synoptic systems on time scales of hours to days, bounded by transient atmospheric fronts that are characterized by strong and intermittent wind stress divergence and convergence (e.g., 3 April 2020 of the animation). 
Latent heat fluxes are also modulated by eddies and fronts, captured by fields of SST and ocean kinetic energy and evolving more slowly over time scales of days to weeks. 
Additionally, southwesterly wind events typically enhance latent heat release above warm eddies while northwesterlies tend to suppress fluxes (e.g., 4 April 2020 vs. 8 April 2020 of the animation). 
Episodic ocean heat loss events associated with southerly cold air outbreaks have also been reported in observations from moorings and uncrewed surface vehicles in the Pacific sector of the Southern Ocean \citep{ogle2018episodic, tamsitt2020mooring, edholm2025synoptic}. In these events, equatorward winds advect cold, dry air over warmer water, increasing air-sea temperature differences and humidity contrast and enhancing the magnitude of latent heat flux, while poleward winds typically bring in warm, moist mid-latitude air over colder SST, in turn weakening the flux with the opposite effect.

\subsection{Distributions of key ocean and atmospheric properties contributing to LHF}\label{disentangling-physical}

\begin{figure}
 \centering
 \noindent\includegraphics[width=.8\linewidth]{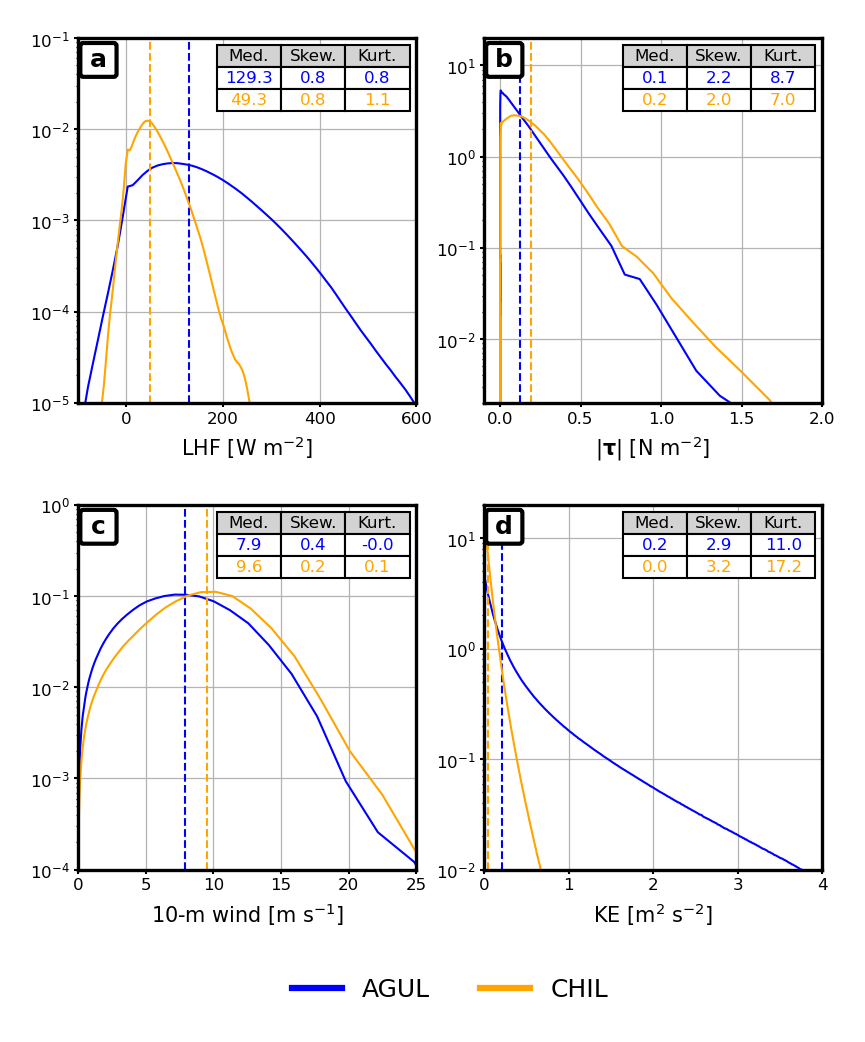}
 \caption{\textbf{Probability density distributions in COAS.} Distributions in AGUL and CHIL regions of (a) latent heat flux, (b) surface wind stress, (c) 10-m wind speed and (d) ocean surface kinetic energy. Distributions are computed over 11 months in COAS. Dashed vertical lines represent the median of each distribution. The legend shows median, skewness and excess kurtosis.}
\label{fig-dist}
\end{figure}

To diagnose the oceanic and atmospheric contributions to LHF variability entering the bulk formula (Eq.~\ref{eq:LHF}), we inspect the distributions of LHF, wind stress, 10-m wind speed and ocean surface kinetic energy (KE).
We compare these distributions across the regions introduced above (Fig.~\ref{fig-dist}). Each distribution includes all grid points and time steps over the 11 simulation months within a given region, and thus represents the combined spatial and temporal variability of each field.

The latent heat flux distribution in the Agulhas Retroflection is shifted toward larger values and is broader than in the Chilean region (CHIL in Fig.~\ref{fig-SOlhf}), indicating both greater mean ocean heat loss and more frequent extreme flux events. This is captured by a greater median latent heat flux in AGUL ($\approx$ 129 W m$^{-2}$) than in CHIL ($\approx$ 49 W m$^{-2}$), despite comparable skewness and excess kurtosis (Fig.~\ref{fig-dist}a). Additionally, the surface current field is much more heterogeneous in the AGUL region, as indicated by the distribution of ocean surface kinetic energy (Fig.~\ref{fig-dist}d) which shows stronger median, skewness and excess kurtosis than in the Chilean region. This is consistent with the snapshots (Fig.~\ref{fig-snapshot}) and animations (see supplemental material) which show an ocean surface shaped by energetic eddies and fronts. However, the distributions of 10-m wind speed and wind stress are similar in both regions, with slightly larger median values in the Chilean region (Fig.~\ref{fig-dist}b,c). 
Together, these results suggest that the broader range of latent heat fluxes in the Agulhas Retroflection are driven primarily by oceanic variability rather than by differences in the overlying atmospheric forcing, consistent with the regional means and snapshots discussed above.

\begin{figure}
 \noindent\includegraphics[width=\linewidth]{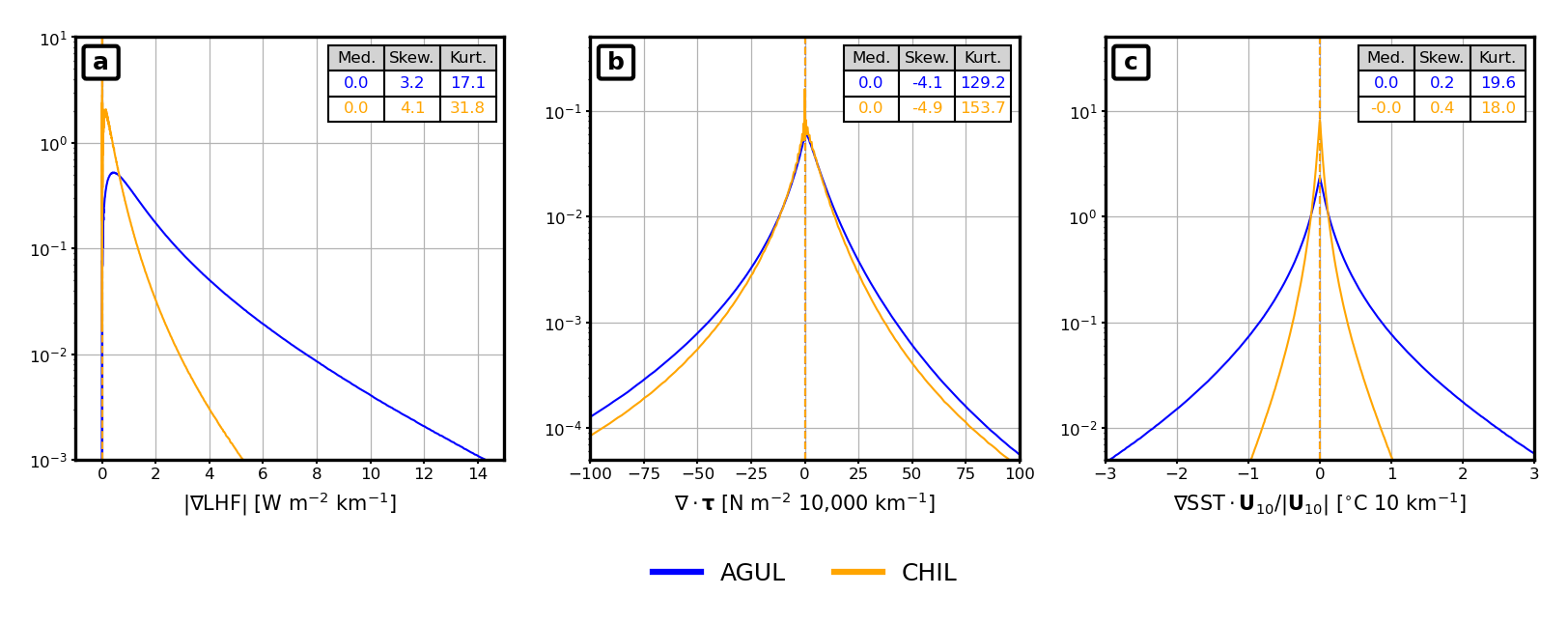}
 \caption{\textbf{Probability density distributions of gradient variables in COAS.} Distributions in AGUL and CHIL regions of (a) the amplitude of the latent heat flux gradient, (b) the divergence of the surface wind stress, (c) the downwind SST gradient. Distributions are computed over 11 months in COAS. Dashed vertical lines represent the median of each distribution. The legend shows the median, skewness and excess kurtosis.}
\label{fig-distgrad}
\end{figure}

The distributions of gradient quantities, namely LHF gradient, surface stress divergence, and downwind SST gradient, provide qualitative insight into finer scale variability (Fig.~\ref{fig-distgrad}). Downwind SST gradients are substantially stronger in the Agulhas Retroflection than in the Chilean region, reaching up to $\pm$ 3 $^\circ$C 10 km$^{-1}$ and $\pm$ 1 $^\circ$C 10 km$^{-1}$, respectively. Given the typical SST ranges in the region, such gradients imply the presence of narrow fronts of order $\mathcal{O}$(1-10km) width, consistent with snapshots (Fig.~\ref{fig-snapshot}). Similarly, extreme LHF gradients reach up to 14 W m$^{-2}$ km$^{-1}$ in the AGUL region, compared with about 5 W m$^{-2}$ km$^{-1}$ off Chile.

However, these distributions alone cannot quantify the dominant temporal scales of latent heat flux variability, nor do they reveal how spatial and temporal scales are coupled. We address this spatiotemporal coupling in the next section using spectral analyses.

\subsection{Disentangling ocean- and atmosphere-driven variability in spectral space} \label{disentangling-spectral}

To characterize the coupled spatial and temporal variability of latent heat fluxes, we use wavenumber-frequency ($k-f$) spectra to partition and quantify the contributions from ocean meso- and submesoscales and from atmospheric synoptic scales and high frequencies, building on the approach of \citet{torres2018partitioning} and \citet{vivant2025ocean} (see methods section \ref{method-spectral-gap}).

\begin{figure}
 \centering
 \noindent\includegraphics[width=.85\linewidth]{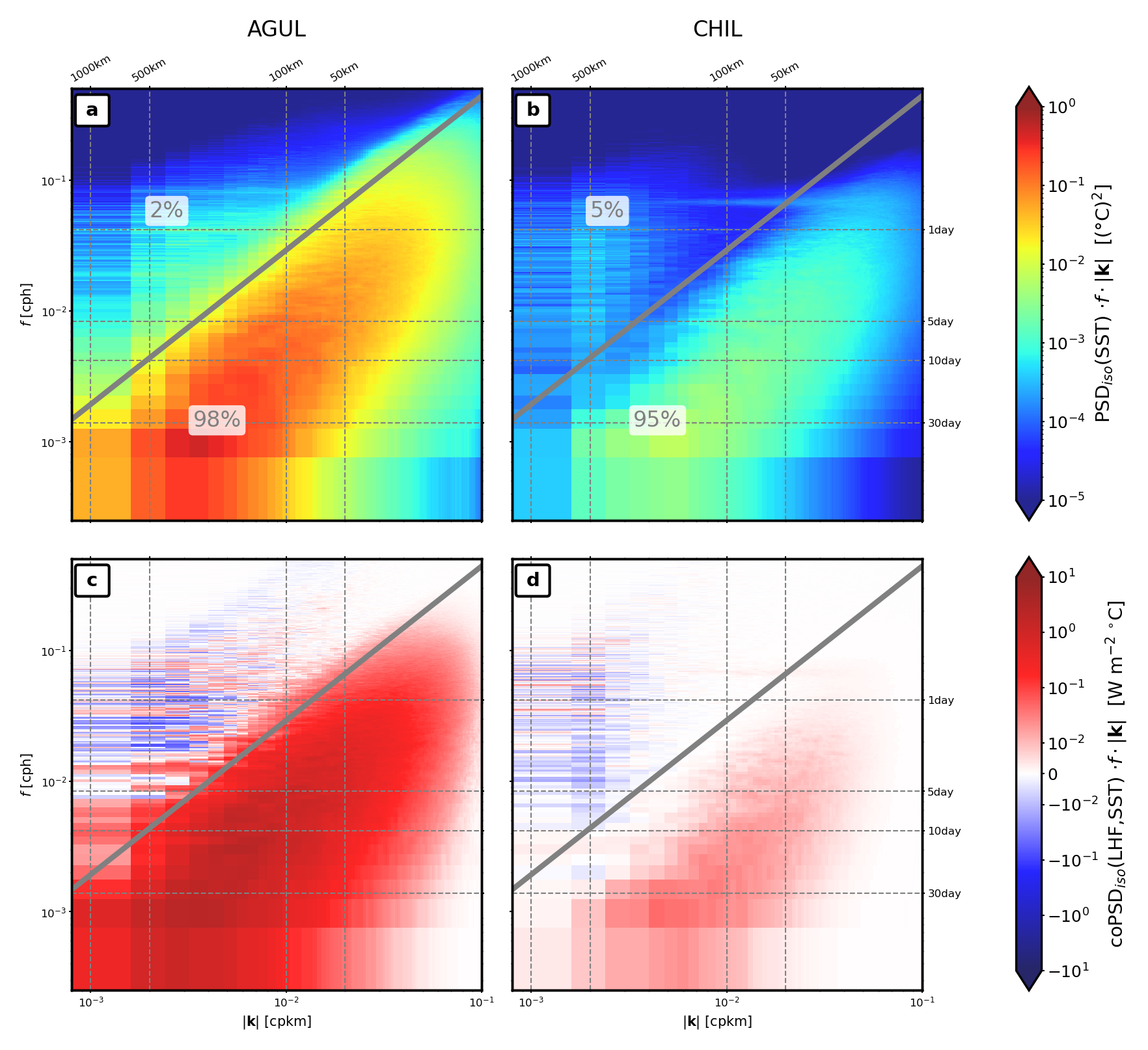}
 \caption{\textbf{SST spectra and SST-LHF cospectra.} Annual-mean, variance-preserving form wavenumber-frequency spectra: (a,b) spectral density of SST, (c,d) cospectral density of LHF and SST. Left column is the AGUL region, right is CHIL. Diagonal partition line as in Fig.~\ref{fig-key}.}
\label{fig-PSDsst}
\end{figure}

\begin{figure}
 \centering
 \noindent\includegraphics[width=.85\linewidth]{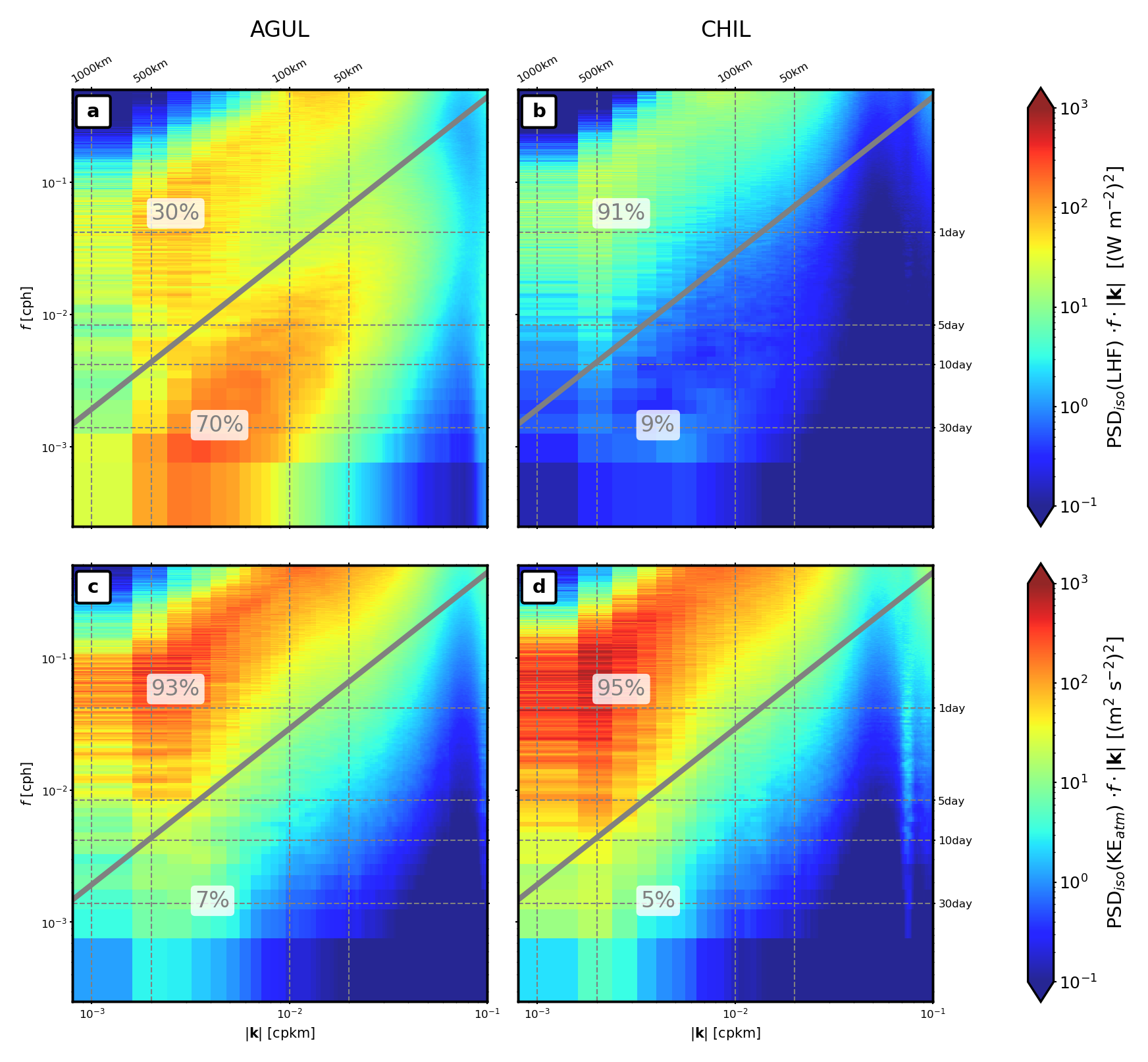}
 \caption{\textbf{LHF and atmospheric kinetic energy spectra.} Annual-mean, variance-preserving form wavenumber-frequency spectra: (a,b) latent heat flux, (c,d) atmospheric kinetic energy at 950 hPa. Left column is the AGUL region, right is CHIL. Diagonal partition line as in Fig.~\ref{fig-key}.
 }
\label{fig-PSDlhf}
\end{figure}

SST variability is concentrated at spatial scales of order $\mathcal{O}$(10-100 km) and temporal scales of order $\mathcal{O}$(weeks), i.e., contained below the diagonal boundary in our spectral partitioning (Fig.~\ref{fig-key}). This accounts for 95\% and 98\% of total SST variance in the CHIL and AGUL regions respectively (Fig.~\ref{fig-PSDsst}a,b). The spectra also highlight the strong regional contrast: the total sea surface temperature variance in AGUL is 56 times larger than in CHIL, consistent with the spatial patterns in physical space (Fig.~\ref{fig-SOlhf}). Additionally, the spectrum of atmospheric kinetic energy aloft confirms that atmospheric variability is primarily contained above the diagonal boundary (Fig.~\ref{fig-PSDlhf}c,d). Together these results support the robustness of our approach for partitioning the ocean and atmosphere contributions to LHF variability.

\clearpage

\begin{figure}
 \centering
 \noindent\includegraphics[width=0.5\linewidth]{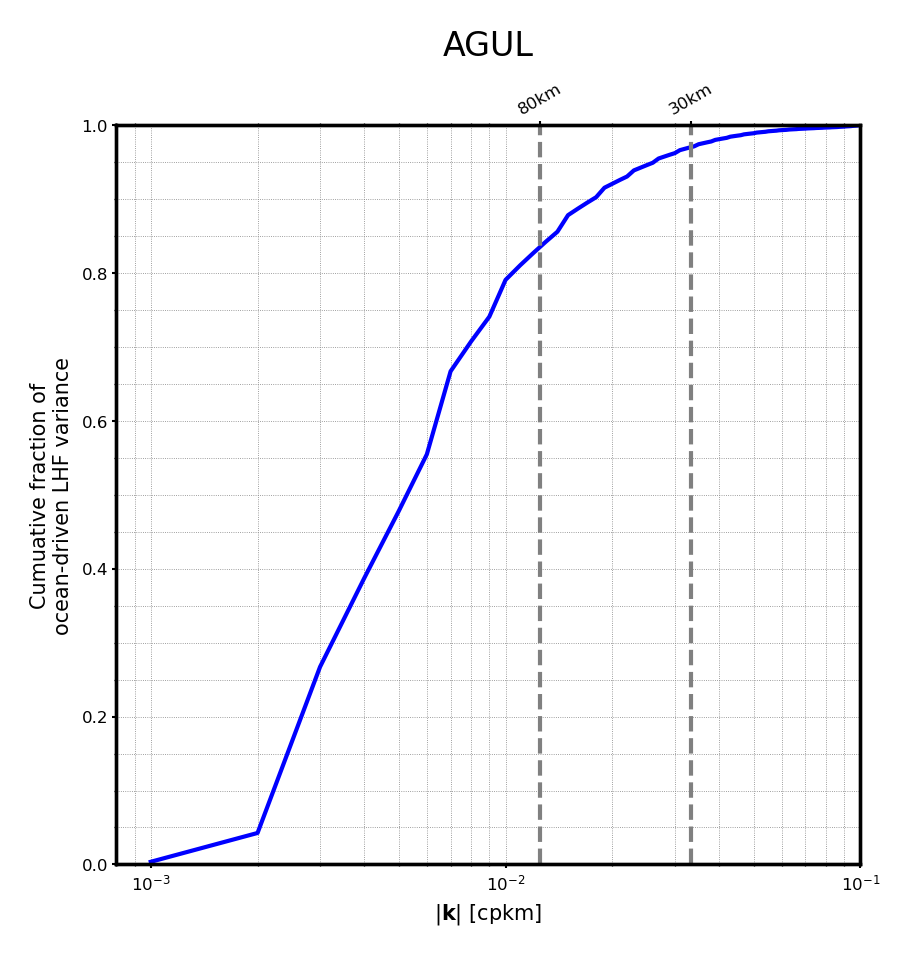}
 \caption{\textbf{Ocean-driven LHF variance explained by meso- to submesoscales.} Cumulative fraction of ocean-driven LHF variance in AGUL, i.e., cumulative integral within the ocean regime below the partition line in Fig.~\ref{fig-PSDlhf}a.}
\label{fig-PSDlhf_cumul}
\end{figure}

The total latent heat flux variance is approximately 2.5 times larger in AGUL than in CHIL. In AGUL, about 70\% of the variance resides in the ocean-dominated sector of the wavenumber–frequency space (i.e., below the diagonal), whereas in CHIL this fraction is only about 9\% (Fig.~\ref{fig-PSDlhf}a,b). To further quantify the role of fine scale processes in the AGUL region, we compute the cumulative fraction of LHF variance below the partition boundary as a function of increasing wavenumber (Fig.~\ref{fig-PSDlhf_cumul}). Defining submesoscales as horizontal scales smaller than the first Rossby radius of deformation, i.e., $\lesssim30$ km in the SO \citep{chelton1998geographical}, we find that they account for $5$\% of the total latent heat flux variance contained in oceanic scales despite being only marginally resolved in COAS. This contribution is therefore likely a lower bound and would be expected to increase in higher-resolution datasets.

\begin{figure}
 \centering
 \noindent\includegraphics{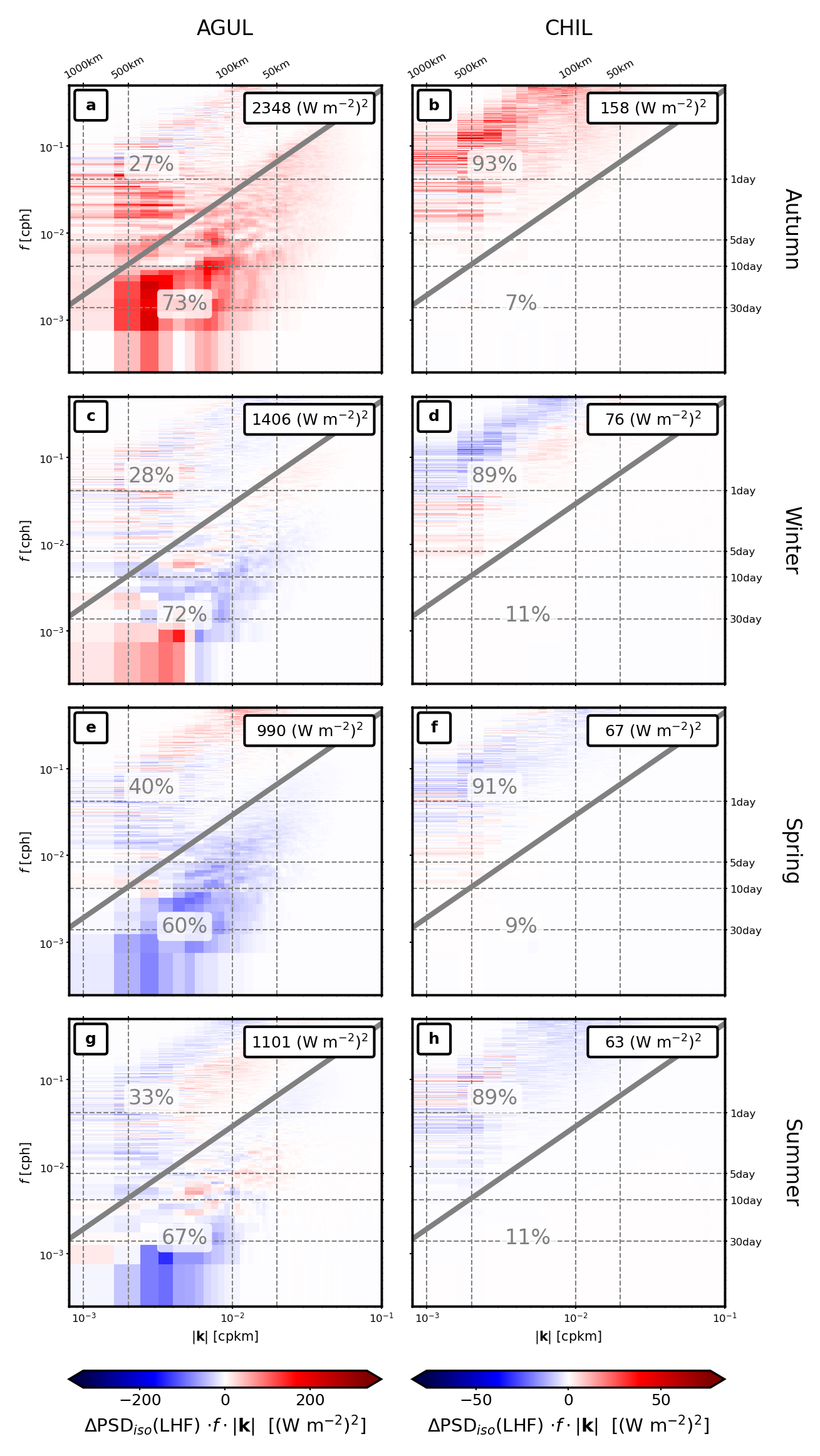}
 \caption{\textbf{Seasonal LHF spectra.} Shading shows seasonal departures of power spectral density from the annual-mean in (a,b) autumn, (c,d) winter, (e,f) spring), and (g,h) summer. For each season, percentages are the relative contribution of each regime to the total LHF variance displayed in the insert in (W/m$^2$)$^2$. Left column is the AGUL region, right is CHIL. Diagonal partition line as in Fig.~\ref{fig-key}.}
\label{fig-seasonlhf}
\end{figure}

Ocean submesoscale activity \citep{sasaki2014impact} and storm tracks \citep{hoskins2005new,nakamura2004seasonal} are known to vary seasonally, motivating an assessment of seasonal changes in LHF variance and its distribution in spectral space. Although the percentage of LHF variance attributed to ocean meso- and submesoscales varies across seasons, the dominant regime in each region remains unchanged: AGUL is persistently ocean-dominated, whereas CHIL remains atmosphere-dominated (Fig.~\ref{fig-seasonlhf}). In the AGUL region, the ocean-driven contribution peaks in the fall (73\%) and reaches a minimum in the spring (60\%). In both regions, the total LHF variance is largest in the fall and weakest in the summer and spring. Seasonal variations in storm activity are known to modulate turbulent heat flux variability \citep{papritz2015climatology, ogle2018episodic, tamsitt2020mooring} and may contribute to seasonal differences in the relative role of atmospheric forcing. However, our limited duration of the simulation (11 months) prevents us from investigating the mechanisms underlying these seasonal differences.

 \clearpage
 \clearpage

In AGUL, the strongly positive co-spectrum between LHF and SST at ocean mesoscales and submesoscales demonstrates that the ocean forces the atmosphere at meso- and submesoscales (Fig.~\ref{fig-PSDsst}c,d), consistent with previous studies showing that SST anomalies at mesoscale induce LHF anomalies of the same sign. Conversely, the co-spectrum is predominantly negative at synoptic and higher-frequency atmospheric scales, consistent with wind-driven cooling events in which enhanced latent heat loss coincides with negative sea surface temperature anomalies \citep{seo2023ocean,xie2004satellite}. Our results extend these findings to fine spatial scales of order $\mathcal{O}$(km), consistent with \citet{strobach2022local} in the Gulf Stream. 

We next investigate patterns of LHF variability across the entire SO (Fig.~\ref{fig-SOlhf}) by extending the spectral partitioning analysis discussed above to the full domain, i.e., 30°S–80°S, 180°W–180°E. At each grid point, we compute the annual-mean LHF spectrum within a region of $\sim\text{1100 km}\times\text{1100 km}$ and estimate the fraction of variance contained in ocean scales, amounting to a total of 13,680 spectra (see methods section~\ref{method-spectral-analysis} for more details).

\begin{figure}
 \centering
 \noindent\includegraphics[width=\linewidth]{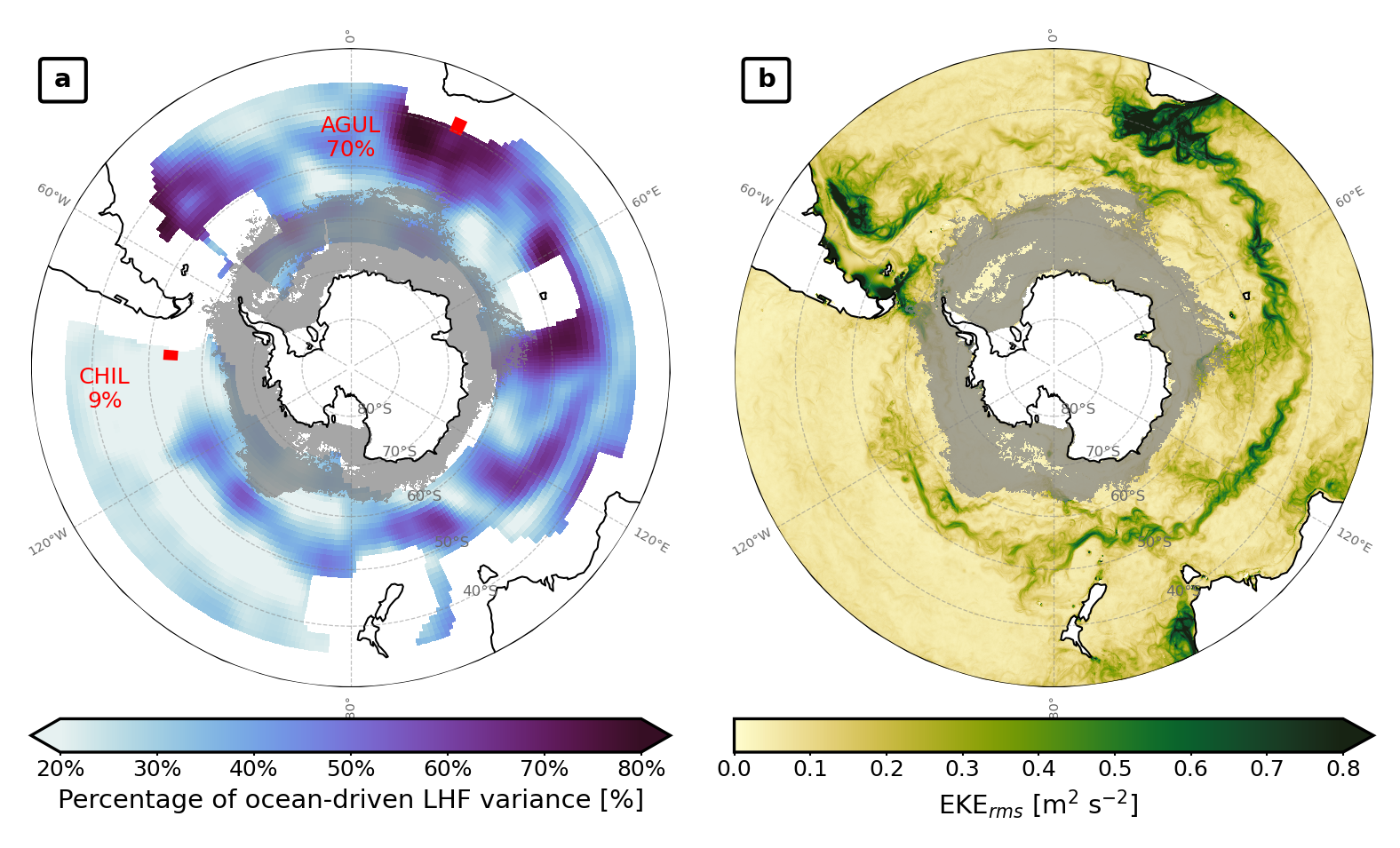}
 \caption{\textbf{Meso- and submesoscale driven LHF variability.} (a) Percentage of ocean-driven LHF variance in $\sim$1100 km $\times \sim$1100 km tiles every 1$^\circ\times$1$^\circ$. For comparison, the percentages derived from AGUL and CHIL regions (Fig.~\ref{fig-PSDlhf}) are in red. (b) Annual root mean square of eddy kinetic energy at the ocean surface. Gray shaded area is where sea ice coverage exceeds 50\%.}
\label{fig-mosaic-map-dlat10}
\end{figure}

Across the Southern Ocean, regions of large ocean-driven latent heat flux variance closely align with regions of elevated annual-mean eddy kinetic energy (Fig.~\ref{fig-mosaic-map-dlat10}). Both are intensified along the ACC, around the Kerguelen Plateau, in the Agulhas Retroflection and Return Current and in the Brazil-Malvinas Confluence. These hotspots also coincide with enhanced SST gradient variability (Fig.~\ref{fig-SOlhf}b), consistent with our findings in the AGUL and CHIL regions. 

\begin{figure}
 \centering
 \noindent\includegraphics[width=.85\linewidth]{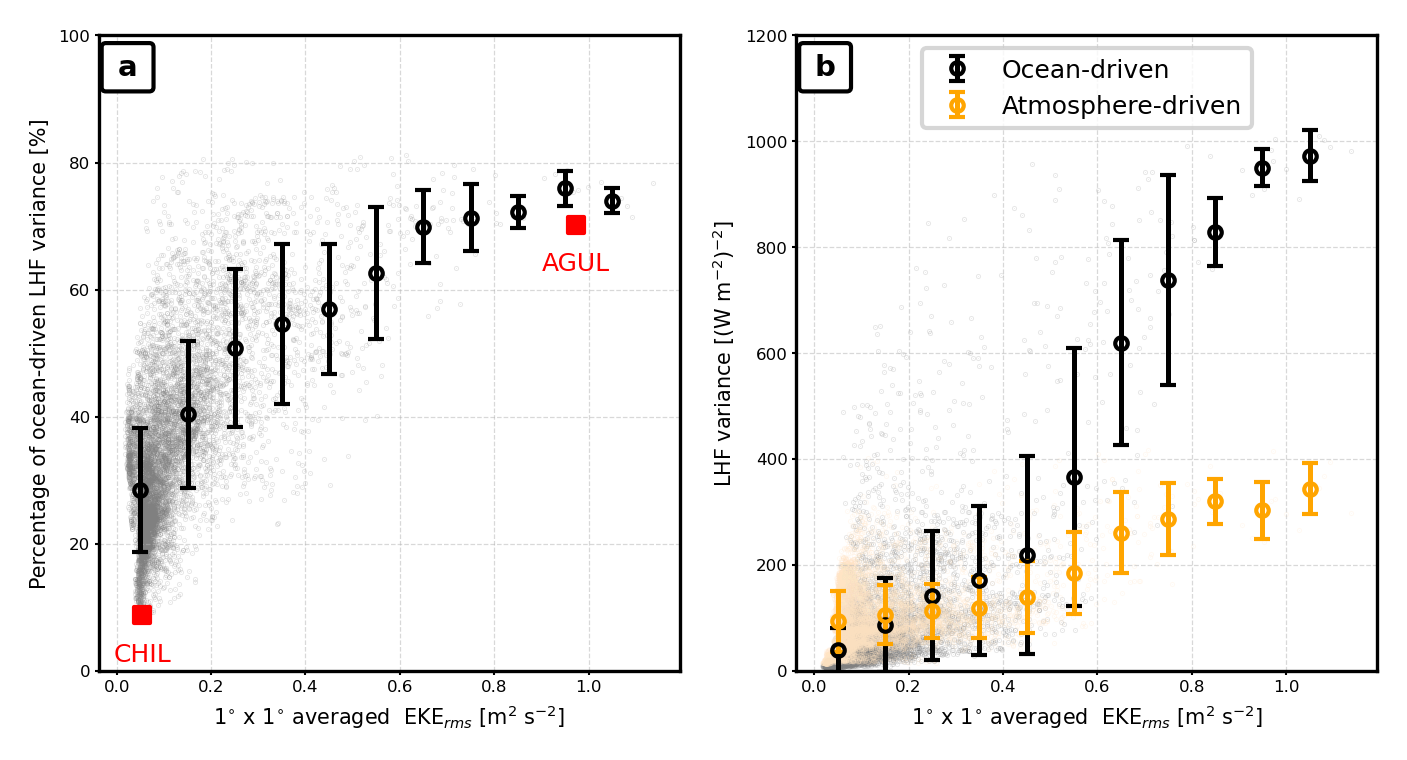}
 \caption{\textbf{Correlation of EKE and LHF variance.} Binned scatter plots of mean EKE against (a) relative ocean-driven LHF variance and (b) total magnitudes of ocean- and atmosphere-driven LHF variance. Scattered dots in the background represent each 1$^\circ\times$1$^\circ$ point in Fig.~\ref{fig-mosaic-map-dlat10}a. Circles and error bars represent the mean and standard deviation inside each EKE bin with at least 3 data points. Percentages from AGUL and CHIL regions are in red. EKE (Fig.~\ref{fig-mosaic-map-dlat10}b) is averaged in each 1$^\circ\times$1$^\circ$ grid cell.}
\label{fig-KE-scatter-dlat10}
\end{figure}

\begin{figure}
 \centering
 \noindent\includegraphics[width=.85\linewidth]{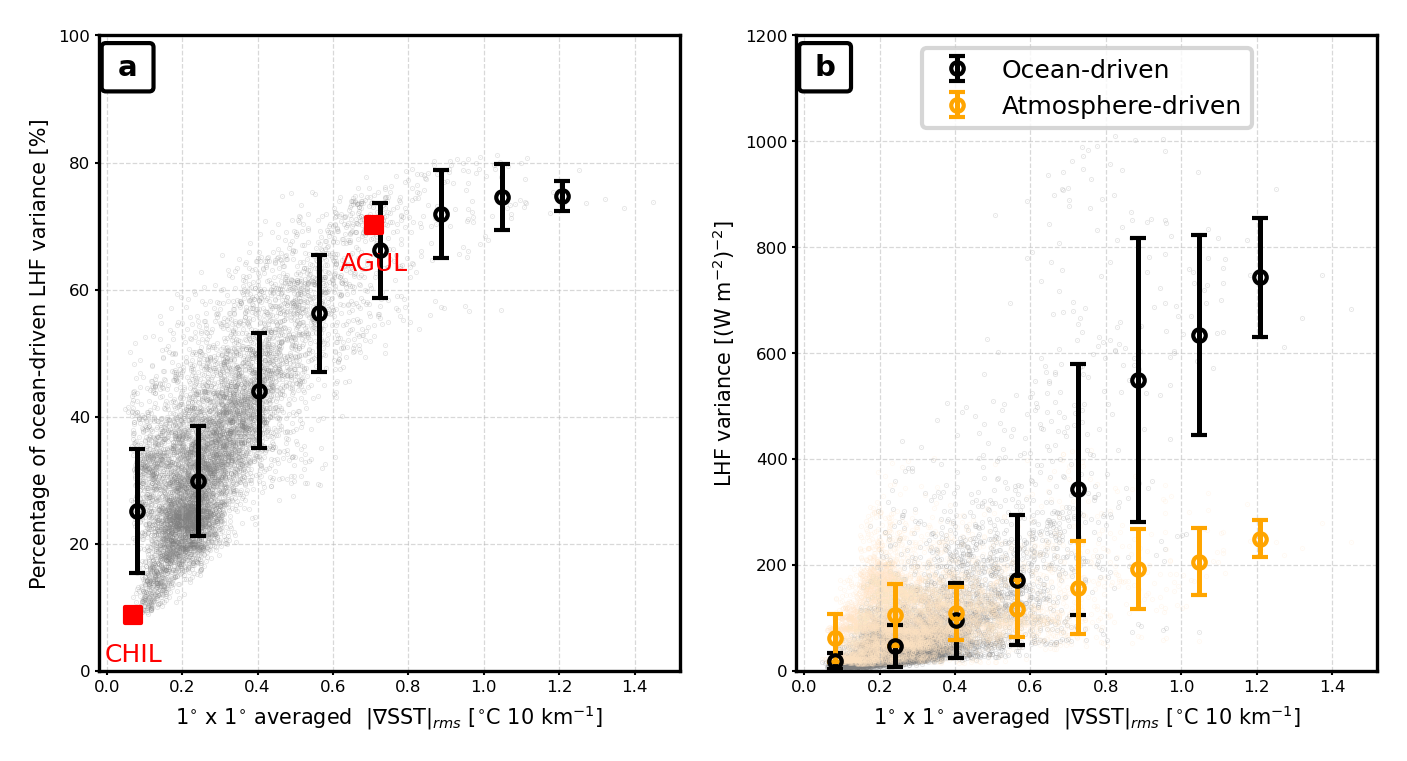}
 \caption{\textbf{Correlation of SST gradient and LHF variance.} Same as Fig.~\ref{fig-KE-scatter-dlat10} but with the root mean square of the SST gradient amplitude (Fig.~\ref{fig-SOlhf}c) on the horizontal axis.}
\label{fig-GRADSST-scatter-dlat10}
\end{figure}

Figure~\ref{fig-KE-scatter-dlat10} reveals that EKE is indeed positively correlated with ocean-driven LHF variance across the entire domain, whether expressed as a fraction of the total variance (in \%, Fig.~\ref{fig-KE-scatter-dlat10}a) or in absolute units (in (W m$^{-2}$)$^2$, Fig.~\ref{fig-KE-scatter-dlat10}b). In eddy-rich regions such as AGUL, the ocean accounts for up to 70-80\% of the total latent heat flux variance. The variability of the SST gradient is also positively correlated with ocean-driven LHF variance (Fig.~\ref{fig-GRADSST-scatter-dlat10}).

\clearpage

The \textit{relative} contribution from ocean processes approaches a plateau at large eddy kinetic energy and strong temperature-gradient variability, implying a persistent atmospheric contribution of at least $\sim$ 20\% regardless of the total LHF variance magnitude (Fig.~\ref{fig-KE-scatter-dlat10}). 
The absolute LHF variance at ocean meso- and submesoscales varies dramatically across the domain (Fig.~\ref{fig-KE-scatter-dlat10}b), by more than two orders of magnitude between the lowest ($\sim$5 (W m$^{-2}$)$^2$) and highest ($\sim$1010 (W m$^{-2}$)$^2$) values. Meanwhile, the atmosphere-driven contribution is much weaker and more homogeneous, spanning from $\sim$13 (W m$^{-2}$)$^2$ to $\sim$417 (W m$^{-2}$)$^2$. Together, these results indicate that much of the pronounced spatial heterogeneity in Southern Ocean latent heat flux variability arises from ocean mesoscale and submesoscale variability rather than from variations in atmospheric forcing\footnote{Note, the variance discussed in this section can not be directly compared to the one shown in Figure~\ref{fig-SOlhf}b, owing to differences in windowing and detrending applied before computing the spectra, see methods section \ref{method-spectral-analysis}.}.

\subsection{Submesoscale-driven local atmospheric response} \label{sms-fronts}

Although submesoscales contribute only $5$\% of the total ocean-driven LHF variance in COAS (Fig.~\ref{fig-PSDlhf_cumul}), previous studies have shown that they can induce a significant local atmospheric response \citep{strobach2022local, vivant2025ocean}. We therefore investigate the effect of submesoscale fronts in the Agulhas Retroflection and Return Current, focusing on latent heat flux gradient, wind stress divergence, convection, and vertical motions.

To isolate the persisting imprint of oceanic features on air-sea fluxes, we apply a centered 5-day rolling window mean to all fields. This low-pass filter removes most of the atmospheric variability, i.e., the upper portion of the spectrum (cf. Fig.~\ref{fig-PSDlhf}) while preserving the mesoscale and submesoscale ocean structure. A sensitivity analysis using averaging windows of 1 to 7 days yields similar results, with increasing noise as the window is made shorter (not shown). We therefore retain the 5-day window as a compromise that preserves strong sea surface temperature gradients while suppressing atmospheric high-frequency variability.

\begin{figure}
 \noindent\includegraphics[width=\linewidth]{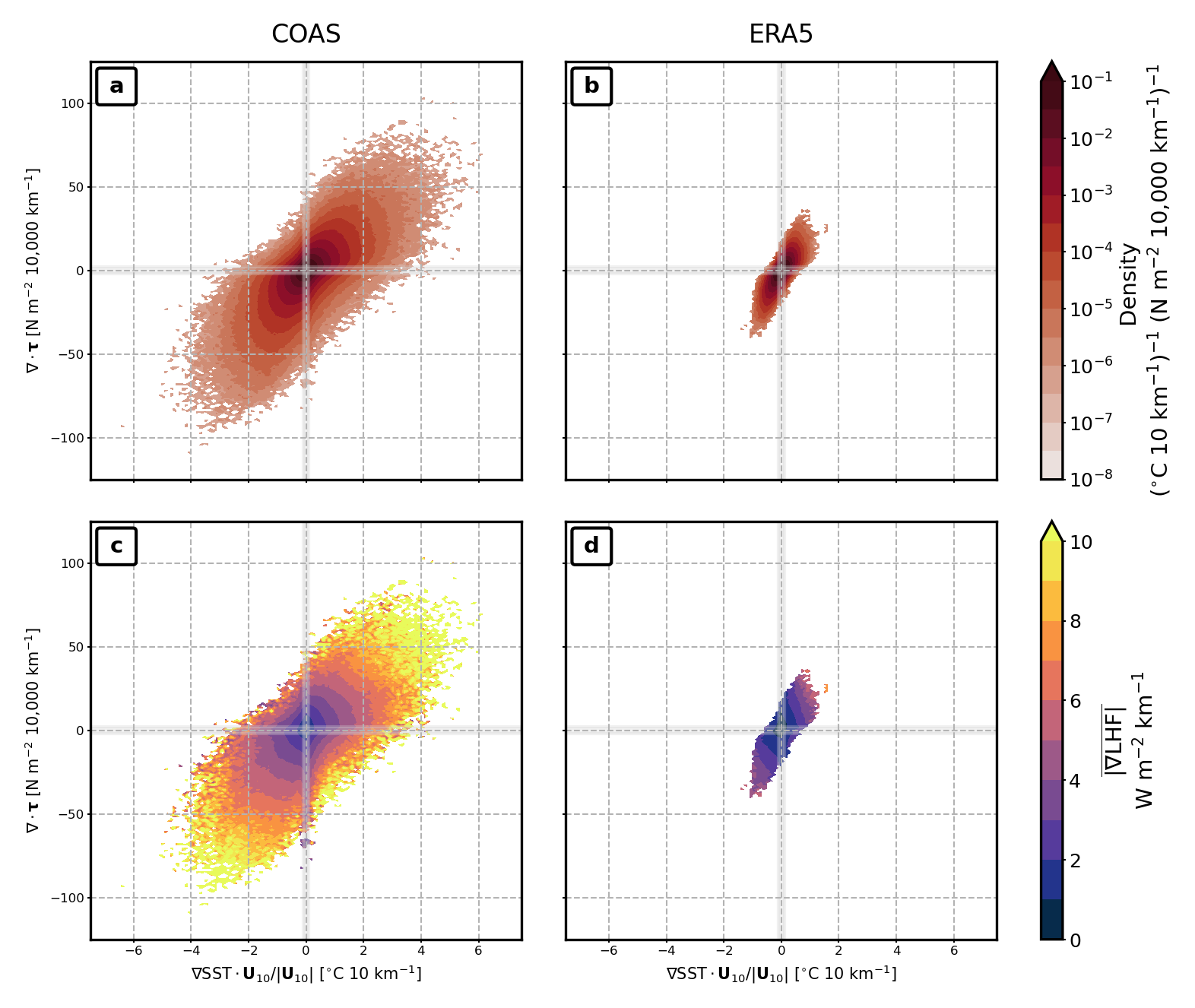}
 \caption{\textbf{Joint distributions in AGUL.} (a,b) Joint probability distribution function of downwind SST gradient and surface stress divergence and (c,d) conditional mean value of LHF gradient. Left is in COAS and right in ERA5. Thick gray lines correspond to the range examined by \citet{chelton2004satellite}.}
\label{fig-jPDF}
\end{figure}

The joint probability density function of wind stress divergence and downwind SST gradient in AGUL (Fig.~\ref{fig-jPDF}a) shows a positive correlation, consistent with \citet{strobach2022local} who used COAS in the Gulf Stream. Given the magnitude of the most extreme 5-day averaged SST gradients, reaching up to $\pm$ 5 $^\circ$C 10 km$^{-1}$, these correspond to submesoscale fronts of order $\mathcal{O}$(1-10 km). Stronger SST gradients have been reported in observations, reaching 24 $^\circ$C 10 km$^{-1}$ in the Kuroshio \citep{yang2024observations} and 15 $^\circ$C 10 km$^{-1}$ in the South Pacific \citep{edholm2025synoptic} across km-scale fronts, suggesting that COAS resolution and the temporal averaging applied here may lead to conservative estimates of the impact of submesoscales. The conditional mean LHF gradient (Fig.~\ref{fig-jPDF}c) increases with the wind stress divergence and the downwind SST gradient. This confirms and generalizes the observations made in the snapshots (Fig.~\ref{fig-snapshot}) and animations (see supplemental material), where we find that the strong LHF and SST gradients are collocated along narrow fronts on the edges of mesoscale eddies. 

Idealized studies have shown that two mechanisms can govern the response of the atmospheric boundary layer in the presence of SST gradients: pressure adjustment and downward momentum mixing (DMM) \citep{foussard2019response}. DMM is associated with the modulation of the vertical shear profile, stability and turbulent mixing in the boundary layer by the underlying SST, transferring more (less) wind momentum downwards over warm (cold) anomalies, and is characterized by a positive correlation between the divergence of wind stress and the downwind SST gradient. Our results reveal that DMM operates down to the submesoscale, consistent with satellite observations in the Gulf Stream and Kuroshio extension \citep{kaouah2025submesoscale}.

To illustrate the dynamics of the local atmospheric response to a submesoscale SST front, we analyze a case study in the Agulhas Retroflection and Return Current. While surface winds predominantly blow West to East across the SO, intermittent northerly synoptic events bring warmer, moister air above relatively colder water, exacerbating the air-sea temperature difference. Wind blowing across a SST front from warm to cool water, e.g., off the edge of an anticylonic eddy, induces wind stress convergence at the front, which can drive collocated upward vertical motion \citep{vivant2025ocean}. Steeper SST gradients, larger SST anomalies and stronger winds are expected to increase the vertical extent of this transport. We select here a case study of northerly winds between July 2 and July 6 of the simulation (see the animation in the supplemental material). As before, we perform a 5-day moving average of oceanic and atmospheric fields to remove the signature of atmospheric high frequencies.

\begin{figure}
 \noindent\includegraphics[width=\linewidth]{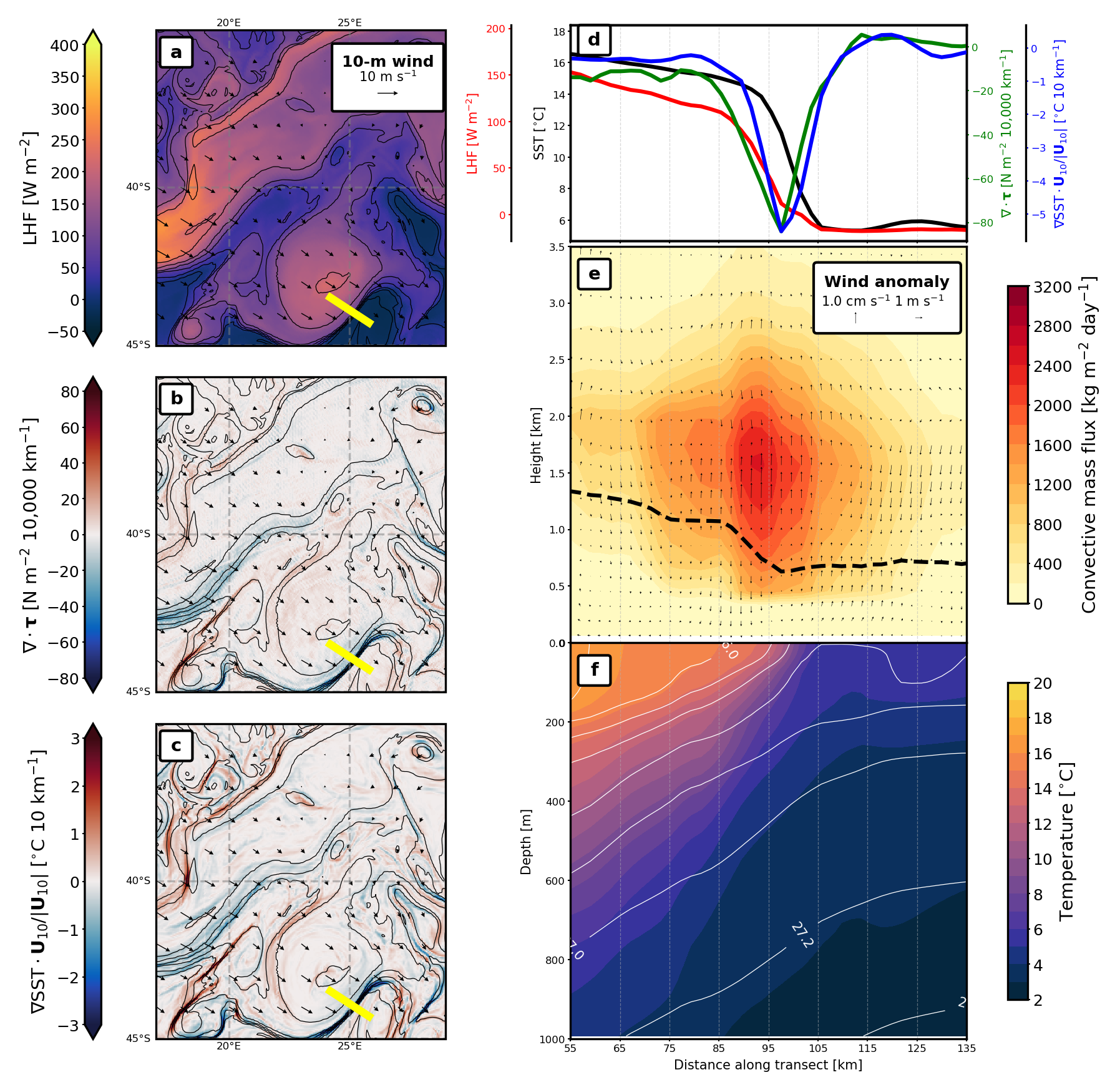}
 \caption{\textbf{Front-induced convection in COAS.} All fields are averaged over five days of northerly winds (July 2 to July 6). Heatmaps of (a) latent heat flux, (b) wind stress divergence, (c) downwind SST gradient. In (a-c), black contours are SST, arrows are 10-m winds, and the thick yellow segment represents the location of the section in (d-f). (d) Along-section SST, LHF, wind stress divergence and downwind SST gradient. (e) Convective mass flux (shading), planetary boundary layer height (dashed line) and along-section wind anomaly (arrows). (f) Temperature (shading) and $\sigma_0$ isopycnals (white contours).}
\label{fig-section_COAS}
\end{figure}

Mesoscale eddies leave a persisting imprint in the the 5-day averaged latent heat flux field, as do the submesoscale fronts in the LHF gradient, wind stress divergence and downwind SST gradient (Figs.~\ref{fig-section_COAS}a,b,c). Winds blowing from the northwest cross a steep SST front on the edge of a warm-core eddy that is centered at about 24$^\circ$E, 44$^\circ$S and about 600-1000 m deep (Fig.~\ref{fig-section_COAS}a,f). The downwind SST gradient and wind stress convergence intensify over the front where they reach up to 80 N m$^{-2}$ 10,000 km$^{-1}$ and -5 $^\circ$C 10 km$^{-1}$ (Fig.~\ref{fig-section_COAS}b,c,d), representative of some of the strongest fronts in the region (Fig.~\ref{fig-jPDF}). 

A convective mass flux is triggered immediately above the front, extending from roughly 500 m to 3500 m above sea level (Fig.~\ref{fig-section_COAS}e). It extends above the planetary boundary layer, whose depth decreases from about 1500 m on the warm side to approximately 600 m on the cold side. The convective mass flux attains its maximum just above the boundary layer, exceeding 2400 kg m$^{-2}$ day$^{-1}$ near 1600 m. Animations (see supplemental material) on two model levels, sitting at roughly 585 m and 1130 m above sea level, show that strong negative downwind SST gradients at submesoscale fronts trigger convective mass fluxes and that they are intensified at the higher level. A previous study using COAS in the Kuroshio Extension showed a similar response above a submesoscale front during a warm wind event, with a locally intensified convective mass flux and secondary circulation extending up to 4 km within the troposphere \citep{vivant2025ocean}. \citet{minobe2008influence} also found that surface wind convergence across the Gulf Stream front triggers vertical velocity anomalies reaching to the upper troposphere and strongest just above the atmospheric boundary layer.

Our results suggest that front-induced convection on the submesoscale is significant in the Agulhas Retroflection and Return Current. Submesoscale SST fronts of order $\mathcal{O}$(km) wide can trigger a local convective mass flux that extends vertically through the boundary layer and horizontally on a narrow scale comparable to the width of the front, contributing to mass and moisture export to the free troposphere.

\subsection{Comparison with ERA5} \label{ERA5}

In the final section of this study, we investigate the impact of omitting fine ocean scales on latent heat flux variability by comparing COAS with ten years of ERA5 reanalysis (2014 – 2024), a product that does not resolve such scales. 
Because a one-to-one comparison cannot be made between a free-running simulation and a reanalysis, the comparison is made in a statistical and phenomenological framework. 

\begin{figure}
 \noindent\includegraphics[width=\linewidth]{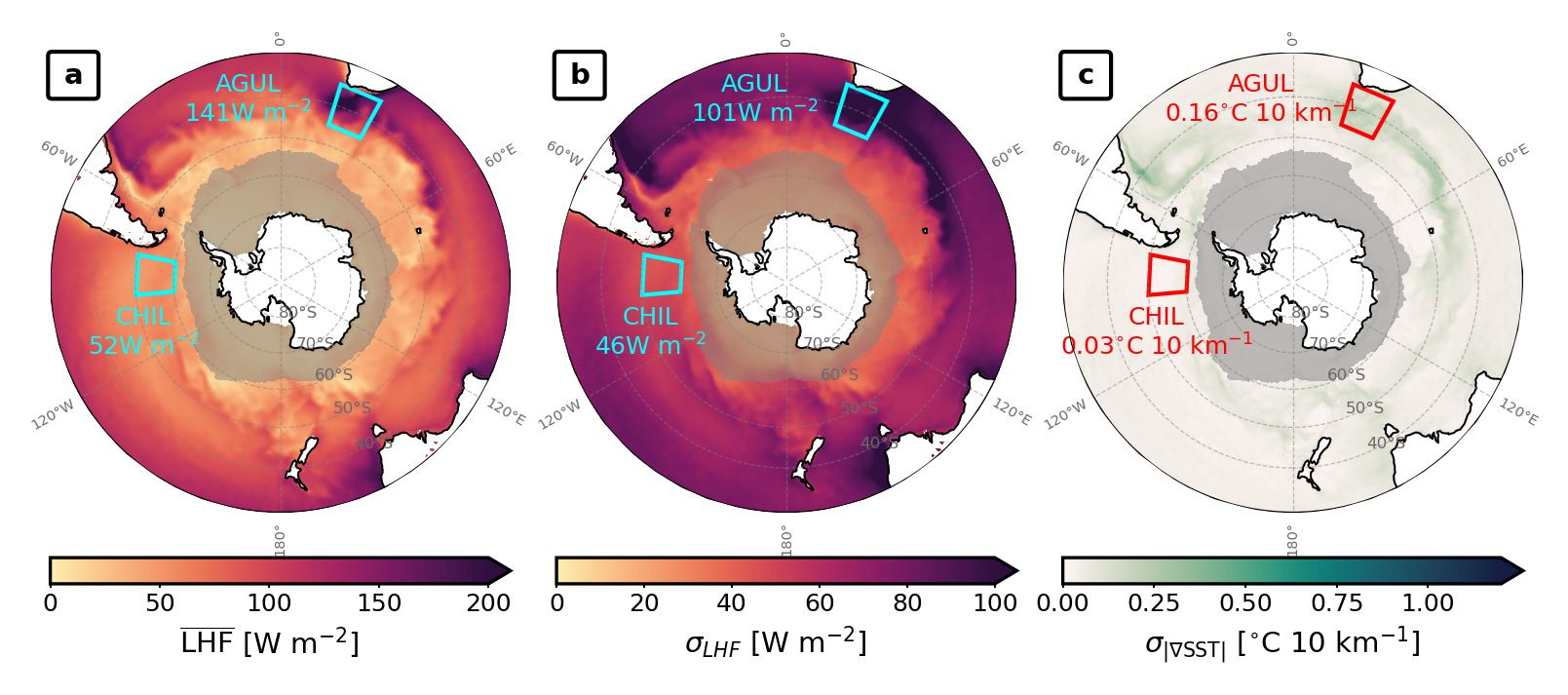}
 \caption{\textbf{Southern Ocean spatial heterogeneity and temporal variability in ERA5.} Same as Fig.~\ref{fig-SOlhf}, from ERA5 reanalysis.}
\label{fig-SOlhf_ERA}
\end{figure}

Across the SO the mean and standard deviation of latent heat fluxes in ERA5 (Fig.~\ref{fig-SOlhf_ERA}a,b) exhibit patterns and orders of magnitude in good agreement with COAS (Figs.~\ref{fig-SOlhf}a,b), although the reanalysis fields are spatially smoother. 
Regionally averaged values in AGUL and CHIL are also comparable.
For example, in the Agulhas Retroflection, the mean and standard deviation are 141 W m$^{-2}$ and 101 W m$^{-2}$ in ERA5, compared with 143 W m$^{-2}$ and 91 W m$^{-2}$ in COAS, and off the tip of Chile, 52 W m$^{-2}$ and 46 W m$^{-2}$ in the reanalysis, versus 54 W m$^{-2}$ and 37 W m$^{-2}$ in the simulation. The slightly larger standard deviation in the ten-year reanalysis likely reflects interannual variability and low-frequency modes that cannot be sampled by a one-year simulation.
Additionally, reanalyses in this region of the world can be plagued with biases, owing to very sparse observations \citep{prend2025observing, truong2022biases, bourassa2013high,liu2011intercomparisons} and outputs can differ substantially among different reanalysis products. For example, using three reanalysis products, \citet{swart2019constraining} reported decadal mean differences reaching up to 50 W m$^{-2}$ and daily mean differences up to 40 W m$^{-2}$ across broad sectors of the SO. Overall, the differences in latent heat flux mean and variance between ERA5 and COAS fall within typical reanalysis uncertainty, supporting the use of ERA5 as a baseline for comparison.

\begin{figure}
 \centering
 \noindent\includegraphics[width=.8\linewidth]{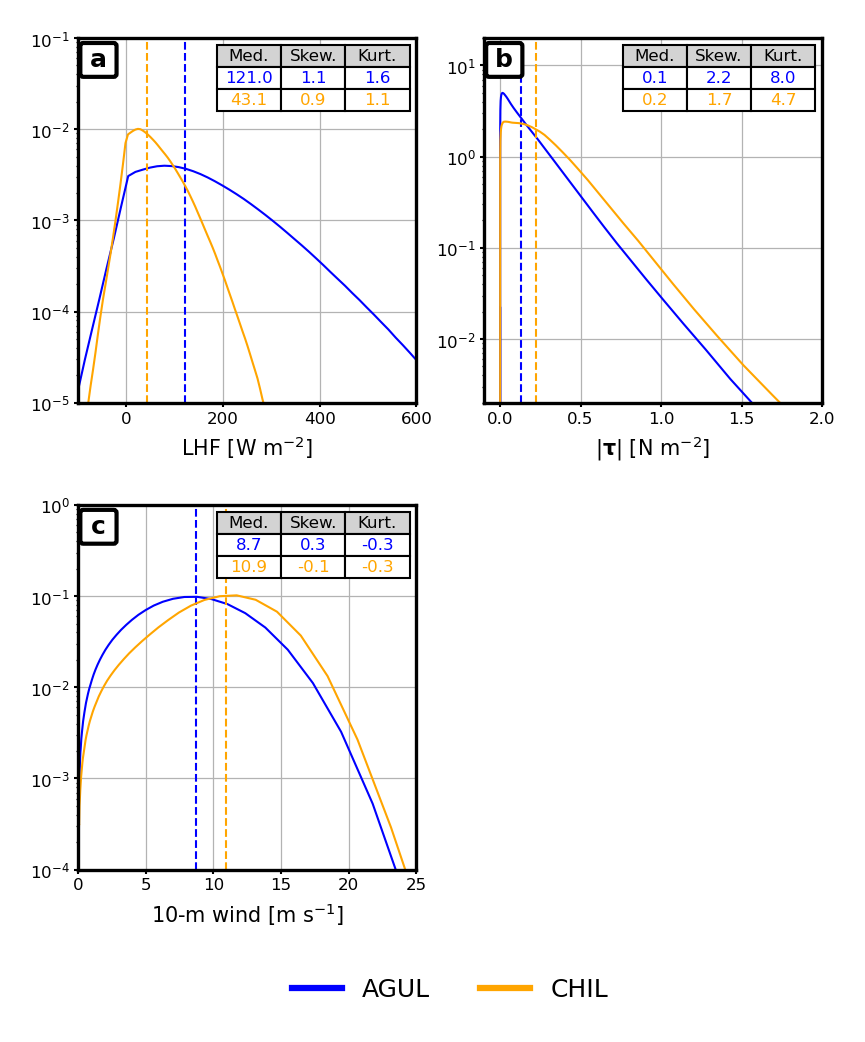}
 \caption{\textbf{Probability density distributions in ERA5.} Same as Fig.~\ref{fig-dist}, from ERA5 reanalysis. Note, ocean kinetic energy is not available in ERA5.}
\label{fig-dist_ERA}
\end{figure}

By contrast, sea surface temperature gradient variability is strongly reduced in ERA5, with on average 0.16 $^\circ$C 10 km$^{-1}$ in AGUL and 0.03 $^\circ$C 10 km$^{-1}$ in CHIL, i.e., about five times smaller than COAS. The ocean forcing in ERA5 is coarser (0.25$^\circ$ horizontal resolution) than the ocean model in COAS (1/24$^\circ$) which smooths the sea surface temperature field and attenuates sharp fronts. Despite this, the mean (Fig.~\ref{fig-SOlhf_ERA}a), standard deviation (Fig.~\ref{fig-SOlhf_ERA}b) and distribution (Fig.~\ref{fig-dist_ERA}a) of latent heat fluxes remain comparable to COAS, potentially reflecting adjustments within the reanalysis flux formulation, though not explored here.

\begin{figure}
 \noindent\includegraphics[width=\linewidth]{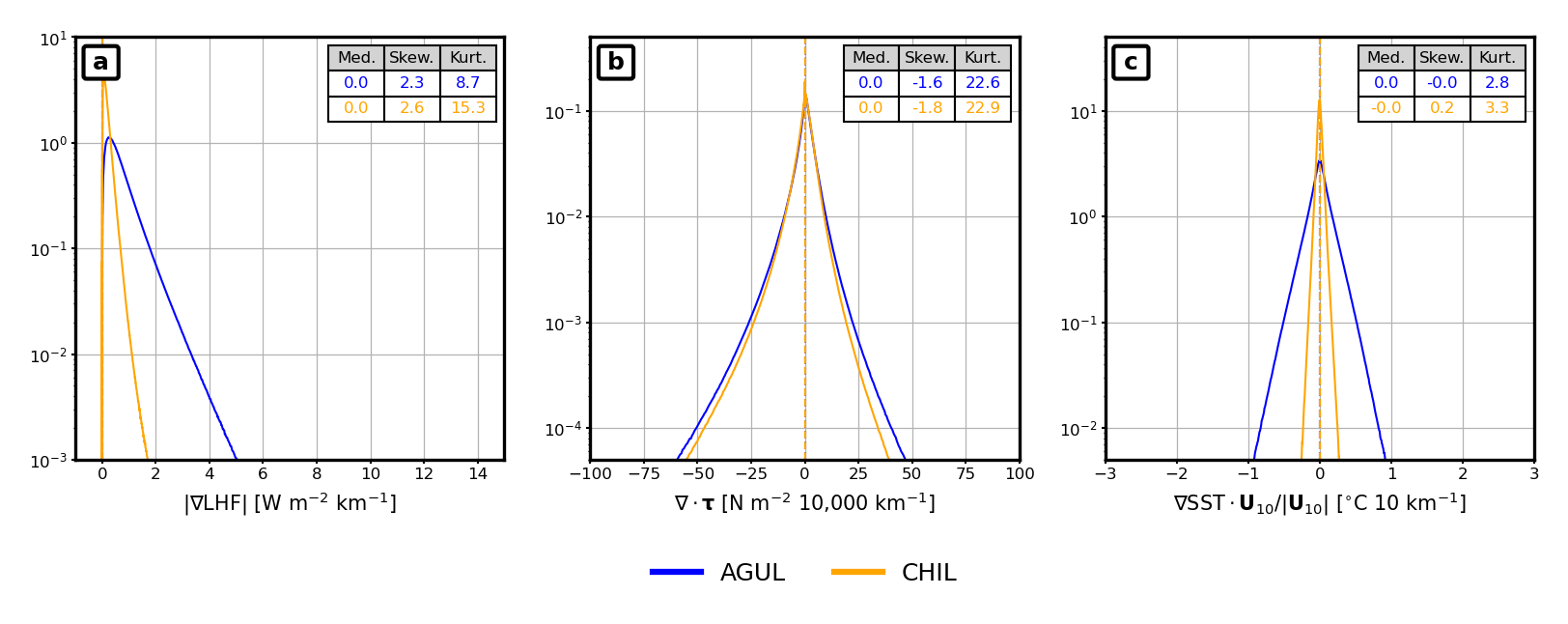}
 \caption{\textbf{Probability density distributions of gradient variables in ERA5.} Same as Fig.~\ref{fig-distgrad}, from ERA5 reanalysis.}
\label{fig-distgrad_ERA}
\end{figure}

The ERA5 distributions of wind stress and 10-m wind speed in AGUL and CHIL regions (Figs.~\ref{fig-dist_ERA}b,c) closely resemble those in COAS (Figs.~\ref{fig-dist}b,c), with similar shape, median, skewness and kurtosis. Note that ocean KE is not available in the reanalysis, only constrained through SST.
In contrast, the distributions of gradient quantities differ substantially between the simulation (Fig.~\ref{fig-distgrad}) and the reanalysis (Fig.~\ref{fig-distgrad_ERA}). Although median values are similar to COAS, the distributions of LHF gradient, surface wind stress divergence, and downwind SST gradient have thinner tails in ERA5, with a kurtosis weaker by a factor of $\sim$2 for $|\nabla \mathrm{LHF}|$, $\sim$4 for $\nabla \cdot \tau$ and $\sim$20 for $\mathbf{k}\cdot\nabla \mathrm{SST}$. This is consistent with the overall smoothing of SST gradients in ERA5 (Fig.~\ref{fig-SOlhf_ERA}c), which suppresses sharp fronts and yields smoother wind-stress divergence. As in COAS, the LHF gradient, downwind SST gradient and wind stress divergence are positively correlated in ERA5, but the joint distributions span a much narrower range in ERA5 (Fig.~\ref{fig-jPDF}b,d) than they do in COAS (Fig.~\ref{fig-jPDF}a,c) further illustrating the absence of extreme gradients associated with submesoscale fronts.

\begin{figure}
 \noindent\includegraphics[width=\linewidth]{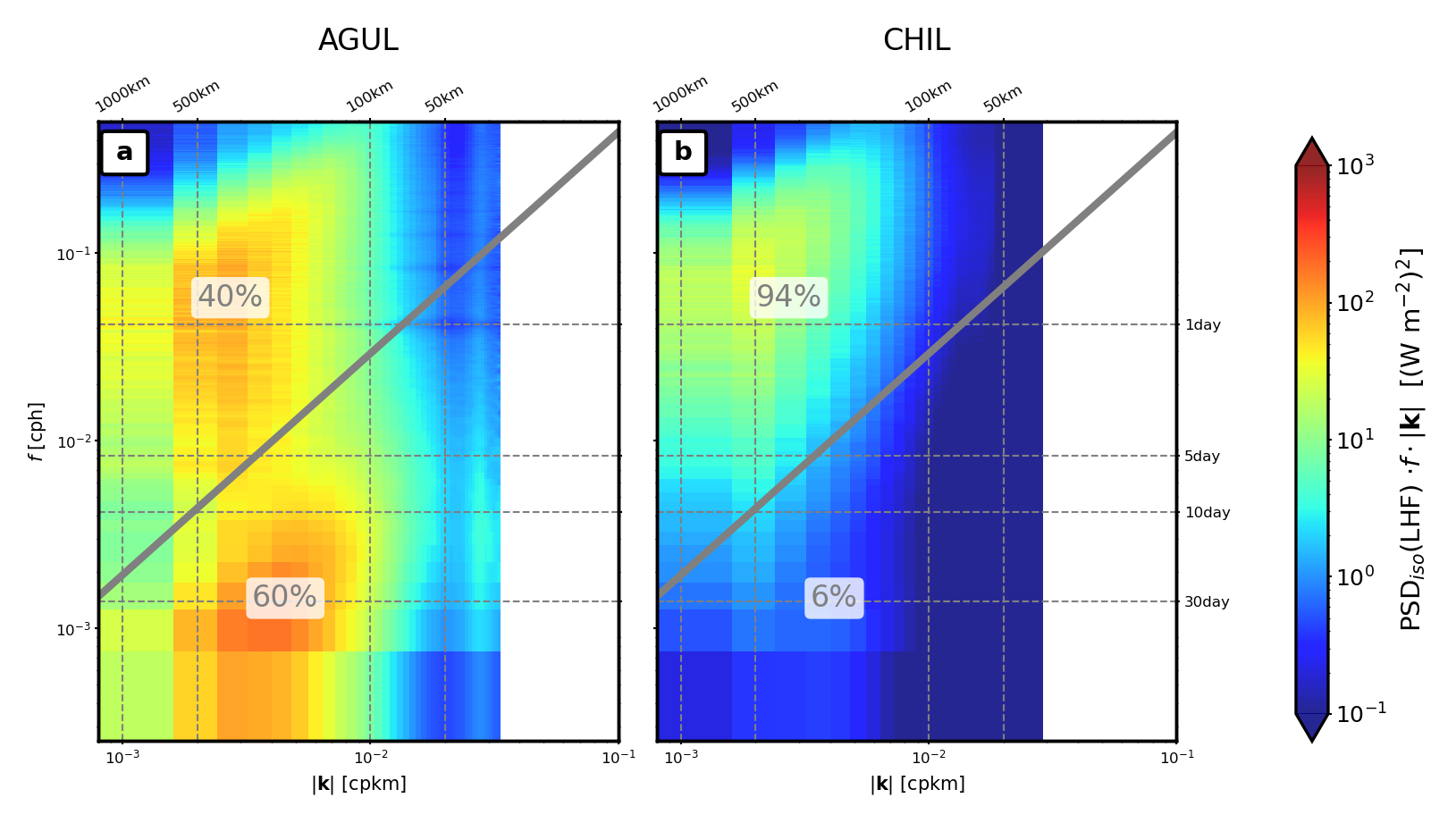}
 \caption{\textbf{LHF spectra in ERA5.} Same as Fig.~\ref{fig-SOlhf}, from ERA5 reanalysis. Note, the Nyquist wavenumber is smaller than in Fig.~\ref{fig-PSDlhf} owing to coarser horizontal resolution in the reanalysis. Dash-dotted line as in Fig.~\ref{fig-key}.}
\label{fig-PSDlhf_ERA}
\end{figure}

In spectral space, ERA5 is limited by a smaller Nyquist wavenumber owing to its coarser horizontal resolution (Fig.~\ref{fig-PSDlhf_ERA}). We define a characteristic scale ($L_c\approx$ 80 km) beyond which the LHF power spectral density rolls off (vertical dash-dotted line in Figures~\ref{fig-key}, \ref{fig-PSDlhf} and \ref{fig-PSDlhf_ERA}). Variability at scales smaller than $L_c$ is therefore poorly captured or entirely missed in ERA5. 
In AGUL, these fine scales contribute $\leq$ 2\% of the ocean-driven LHF variance in the reanalysis (not shown), compared to $\approx$17\% in COAS (Fig.~\ref{fig-PSDlhf_cumul}). While total LHF variance in ERA5 is comparable to COAS, the reanalysis underestimates the fraction of ocean-driven LHF variability, especially in the AGUL region (about 60\% in ERA5 versus 70\% in COAS; Fig.~\ref{fig-PSDlhf_ERA}). Qualitatively, this difference can be explained by a smaller Nyquist wavenumber truncating the wavenumber-frequency space (Fig.~\ref{fig-key}) in such a way that the subset attributed to the oceanic influence (below the partition line) is cropped more than the subset attributed to the atmospheric influence (above the partition line). In other words, coarser resolution attenuates fine scale structures, leading to a more severe underestimation of oceanic variability where small scales are more prevalent than in the atmosphere.

\begin{figure}
 \noindent\includegraphics[width=\linewidth]{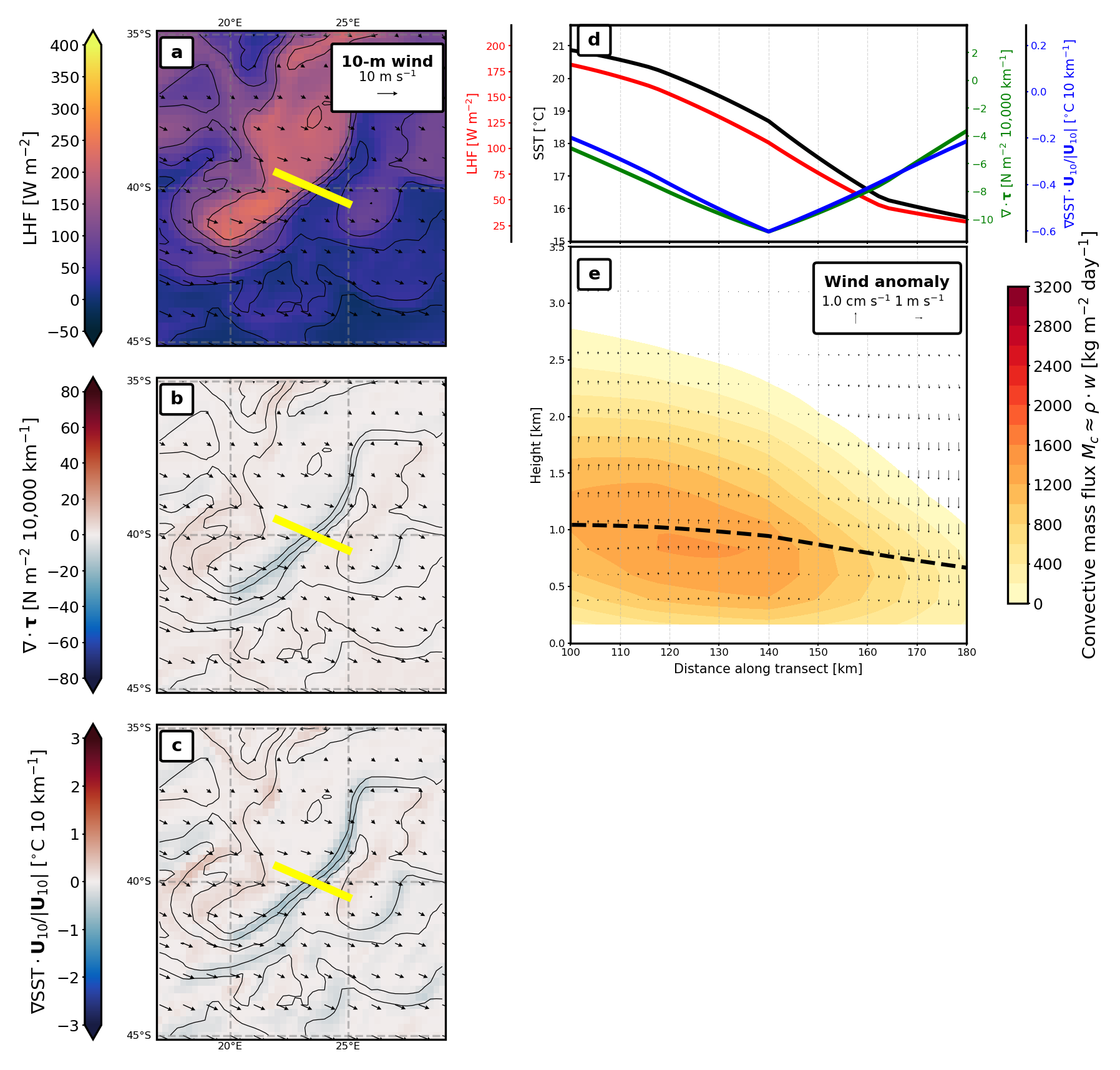}
 \caption{\textbf{Absence of front-induced convection in ERA5.} Same as Fig.~\ref{fig-section_COAS}, from ERA5 reanalysis. Convective mass flux is approximated as $M_c = \rho w$ (see methods). Note: No subsurface oceanic data are available in ERA5; only SST is provided}
\label{fig-section_ERA}
\end{figure}

The same truncation also limits the highest-wavenumber portion of the atmospheric sector, implying that small-scale atmospheric responses are attenuated as well. To illustrate this, we examine a case study between June 9 and 14, 2024, of the reanalysis in the Agulhas Retroflection (Fig.~\ref{fig-section_ERA}). 
As in the COAS case study previously presented, northwesterly winds cross a temperature front near 23$^\circ$E, 40$^\circ$S from warm to cold water. However, the downwind temperature gradient and wind-stress divergence are weaker and spatially broader than in COAS, not exceeding about $-0.6^\circ$C per 10 km and 10 N m$^{-2}$ per 10,000 km (Fig.~\ref{fig-section_ERA}b-d). 
The atmospheric response occurs on the same length scale as the ocean mesoscale, with vertical motions upward (downward) on the warm (cold) side of the front, rather than an intense updraft collocated with the front itself. Consistent with a limited effective resolution (Fig.~\ref{fig-PSDlhf_ERA}), ERA5 does not exhibit an atmospheric response in phase with the sea surface temperature front \citep{vivant2025ocean}, and therefore misses the localized convection and boundary-layer mass export associated with submesoscale coupling.

\clearpage

\section{Conclusion and Discussion}\label{conclusion}

Recent studies show that the variability of air-sea heat fluxes at scales of order $\mathcal{O}$(100 km) is forced by SST anomalies due to mesoscale ocean variability, especially in western boundary currents \citep{small2019air, seo2023ocean}. In this study we expand these results and quantify the relative contribution of oceanic and atmospheric scales to LHF variability across the Southern Ocean, using a global, kilometer-scale coupled ocean–atmosphere simulation run over 11 months. In particular, we investigate the role of mesoscale eddies in driving LHF variability and submesoscale fronts in inducing a local atmospheric response.

Annual-mean latent heat flux in regions of elevated eddy kinetic energy — such as along the Antarctic Circumpolar Current, within the Agulhas Retroflection, and at the Brazil–Malvinas Confluence — reaches up to $\approx$ 215 W m$^{-2}$, roughly three times larger than in low-eddy sectors such as the southeast Pacific. In these eddy-rich regions, ocean mesoscales and submesoscales explain about 70 – 80\% of total LHF variance whereas they contribute less than 10\% in eddy-poor sectors. The contribution from fine ocean scales to LHF variance increases proportionally with eddy kinetic energy and with the variability of sea surface temperature gradients (Fig.~\ref{fig-KE-scatter-dlat10}), while the atmosphere-driven contribution to latent heat flux variance is comparatively weaker and more spatially uniform across the Southern Ocean.

Focusing on a case study in the eddy-rich Agulhas Retroflection and Return Current, we show that downward momentum mixing drives the local atmospheric response to SST anomalies down to submesoscales, consistent with previous findings in the Gulf Stream and Kuroshio Extension \citep{strobach2022local,vivant2025ocean} and in an idealized model \citep{foussard2019response}. 
We further demonstrate that submesocale SST fronts of order 5 $^\circ$C 10 km$^{-1}$ induce convergent wind stress motions, driving a secondary vertical circulation. This secondary circulation extends above the planetary boundary layer, and is associated with an enhanced convective mass flux of up to 2400 kg m$^{-2}$day$^{-1}$ localized directly above the SST front, similar to recent COAS results in the Kuroshio Extension \citep{vivant2025ocean}.

Comparison between COAS and ERA5 shows that variability at spatial scales $\leq$~80 km is poorly represented in ERA5, even though these scales account for roughly 17\% of the ocean-driven latent heat flux variance in COAS. As a result, in the reanalysis we observe a subdued atmospheric response that occurs on the same length scale as the ocean mesoscale anomaly, rather than an intense one right above the sharp SST front, like seen in COAS.

Taken together, these results underscore a tight coupling between fine ocean and atmospheric scales, that remains crucially unresolved in state-of-the-art reanalyses \citep{hersbach2020era5} and current climate models \citep{eyring2016overview, chang2020unprecedented}.

Model resolution remains a limitation of this study: COAS only partially resolves ocean submesoscales at high latitudes, where the first Rossby radius of deformation is $\mathcal{O}(10-30~\text{km})$ \citep{chelton1998geographical}. As a result, our analysis might underestimate the steepness of submesoscale fronts and the strength of the associated surface wind convergence, with potential impacts on the representation of updrafts and extreme precipitation \citep{chang2025future}. Submesoscale contributions to latent heat flux variability may therefore be larger than estimated here.
In addition, we did not examine the influence of seasonal sea-ice cover on latent heat flux variability. In the presence of sea-ice, both the physics and parameterization of air-sea heat exchange (cf. Eq.~\ref{eq:LHF}) are modified \citep{swart2019constraining, elvidge2021surface}. Even partial ice cover may decouple wind stress from the underlying SST features, potentially modifying the local atmospheric response.

\begin{figure}
\centering
 \noindent\includegraphics[width=0.6\linewidth]{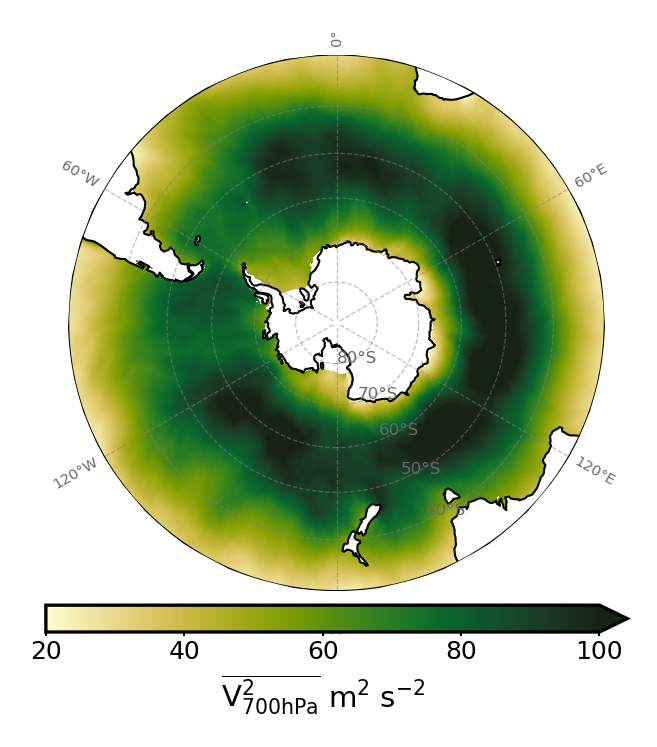}
 \caption{\textbf{Tropospheric meridional kinetic energy.} Annual meridional KE at 700 hPa, i.e. $\overline{\mathrm{V}_{\mathrm{700hPa}}^2}$.}
\label{fig-V700hPa}
\end{figure}

These findings also have implications for large-scale atmospheric circulation and water cycle. Our results imply that ocean meso- and submesoscales have the potential to intensify atmospheric storms by enhancing latent heat release and driving convergent and divergent motions in the lower troposphere \citep{foussard2019storm, vivant2025ocean,vivant2025meandering}. Fine ocean scales may also induce mass export above the marine atmospheric boundary layer and impact remote precipitations associated with atmospheric rivers \citep{maclennan2022contribution, chang2025future, liu2021ocean}. 
In addition, moist processes driven by surface latent heat fluxes have the potential to energize the tropospheric kinetic energy at mesoscale ($\leq$ 500 km) \citep{hamilton2008mesoscale,waite2013mesoscale}. In our simulation, the  annual mean meridional kinetic energy at 700 hPa (Fig.~\ref{fig-V700hPa}) appears spatially homogeneous, in stark contrast with the patchy LHF distribution observed across the Southern Ocean (Fig.~\ref{fig-SOlhf}). Similarly, surface wind-stress values in AGUL and CHIL are comparable despite drastically different LHF distributions in these two regions (Fig.~\ref{fig-dist}). This suggests a non-local atmospheric response to ocean forcing by mesoscale eddies and submesoscale fronts in the Southern Ocean. 
Overall, although unresolved in most climate models and widely used reanalyses, ocean meso- and submesoscales emerge as a potentially important pathway for influencing the atmosphere both locally and beyond the forcing region. 

\newpage

%

\clearpage
\acknowledgments

LR acknowledges support from the physical oceanography program at the Office of Naval Research (ONR). LS and LL acknowledge support from NASA grants 80NSSC24K1653 \& 80NSSC23K0985 and NSF EAGER award 2428537. This work would not have been possible without the COAS simulation made available by Dimitris Menemenlis and Andrea Molod, nor without the help provided by Hector Torres for accessing the data. The authors thank Greg Sinnett for valuable feedback that improved the manuscript.

%
%
\datastatement

All data used in this study are publicly available. Data from the ERA5 reanalysis are available at \url{http://cds.climate.copernicus.eu/}. The outputs of COAS can be found at: \url{https://portal.nccs.nasa.gov/datashare/G5NR/DYAMONDv2/GEOS_6km_Atmosphere-MITgcm_ 4km_Ocean-Coupled/}. Animations of the COAS outputs are readily available at: \url{https://data.nas.nasa.gov/geoseccoviz/geoseccovizdata/c1440_llc2160/GEOS/index.html}. The code for the analysis presented here is available at \url{https://zenodo.org/records/18511904}.


\appendix[A]
\appendixtitle{Power spectral density and cospectrum} 

A discrete 3-D FFT is computed from P hourly snapshots of size N x M, where N and M are the number of zonal and meridional samples. Before taking the FFT of a discrete signal $a(m,n,p)$ its linear trend is removed and a Hanning window is applied along all three dimensions. The detrended, windowed signal is then zero-padded in longitude and latitude.

Consider a discrete, real signal $a(m,n,p)$ and window function $w(m,n,p)$, $ m\in[0..M-1]$, $ n\in[0..N-1]$, $p\in[0..P-1]$. Then its discrete Fourier transform is $\hat{A}(k,l,\omega)$:

\[
\hat{A}(k,l,\omega) =
   \sum_{p=0}^{P-1} \sum_{m=0}^{M-1} \sum_{n=0}^{N-1} 
   w(m,n,p) 
   \, a(m,n,p)
   \, e^{-2\pi i \frac{n k}{N}}
   \, e^{-2\pi i \frac{m l}{M}}
   \, e^{-2\pi i \frac{p \omega}{P}}
\]

where discrete zonal and meridional wavenumbers $k,l$, and frequency $\omega$ are:
\[
\begin{split}
k&=[-N/2, ..., -1, 0, 1, ..., N/2-1] \cdot 1/L_{lon} \nonumber \\
l&=[-M/2, ..., -1, 0, 1, ..., M/2-1] \cdot 1/L_{lat} \nonumber \\
\omega&=[-P/2, ..., -1, 0, 1, ..., P/2-1] \cdot 1/T \nonumber
\end{split}
\]

Parseval's identity relates the variance of the spectral and physical signals:

\[
\frac{1}{(N M P)^2} \sum_{k,l,\omega} \left| \hat{A}(k,l,\omega) \right|^2 
= \frac{W_{ss}}{N M P} \sum_{m,n,p} \left| a(m,n,p) \right|^2
\label{eq:parseval}
\]

where $W_{ss}$ accounts for the power attenuation that arises from windowing:
\[
W_{ss} = \sum_{m,n,p} \left| w(m,n,p) \, a(m,n,p) \right|^2  / \sum_{m,n,p} \left| a(m,n,p) \right|^2
\]

The power spectrum must be corrected by a factor $\frac{1}{W_{ss}(NMP)^2}$ to recover the original variance. Finally, the power spectral density is obtained as
\[
\left|\hat{a}(k,l,\omega) \right|^2 = \frac{1}{W_{ss}(NMP)^2} \frac{1}{(dk \, dl \, d\omega)}\left| \hat{A}(k,l,\omega) \right|^2 \label{eq:PSDdir}
\]
where spectral resolutions are:

$dk=1/L_{lon}, dl=1/L_{lat}, d\omega=1/T$.

The isotropic power density is obtained by integrating in the directional wavenumber space ($k, l$): 
\[
\left|\hat{a}_{iso}(k_{iso}, \omega) \right|^2= \sum_{(k,l) \in \mathscr{A}(k_{iso})} \left|\hat{a}(k, l, \omega)\right|^2 \; k_{iso} \, d\theta
\]

where:

$\mathscr{A}(k_{iso}) = \Bigl\{k,l \mid \sqrt{k^2+l^2} \in [k_{iso}-\frac{dk_{iso}}{2},k_{iso}+\frac{dk_{iso}}{2}]\Bigr\}$ 

is the ring of radius $k_{iso}$ and width $dk_{iso}$, and $d\theta$ uniformly weights the integral. This assumes that $k,l$ are uniformly distributed. The isotropic wavenumber discretization is arbitrarily chosen, whereas frequency spacing $df=1/T$ and maximum $f_{max}$ result directly from the length $T$ and sampling in each FFT window.

Similarly, for two discrete signals $a(m,n,p)$ and $b(m,n,p)$, with respective spectra $\hat{A}(k,l,\omega)$ and $\hat{B}(k,l,\omega)$, their co spectral density is defined as:

\[
\widehat{ab}(k,l,\omega) =  \frac{1}{W_{ss}^{ab}(NMP)^2} \frac{1}{(dk \, dl \, d\omega)}
   \text{Re}\!\bigl(\hat{A}(k,l,\omega) \, \hat{B}^*(k,l,\omega)\bigr)
   \]

where $W_{ss}^{ab}$ accounts for the power attenuation that arises from windowing: 
\[
W_{ss}^{ab} = 
\frac{ \sum_{m,n,p} \left| w(m,n,p) \, a(m,n,p) \, b(m,n,p) \right|^2 }
     { \sum_{m,n,p} \left| a(m,n,p) \, b(m,n,p) \right|^2 }
\]

\appendix[B]

\appendixtitle{Spectral gap fit}

We define a linear divide through the spectral gap as follows. First, the PSD is interpolated from variance-preserving form onto a regularly spaced $(k-f)$ grid after fitting with a 2D spline. The interpolated PSD is smoothed using a Gaussian kernel before searching for local minima in the spectral gap and fitting a linear regression through them. Owing to the limited degrees of freedom available for spectral averaging (four segments of 83 days), this method picks up a few spurious minima. To ensure a robust fit, the regression parameters are estimated using a random sample consensus algorithm (RANSAC), an iterative procedure for model fitting in the presence of outliers \citep{fischler1981random}. 

\begin{figure}
 \centering
 \noindent\includegraphics[width=\linewidth]{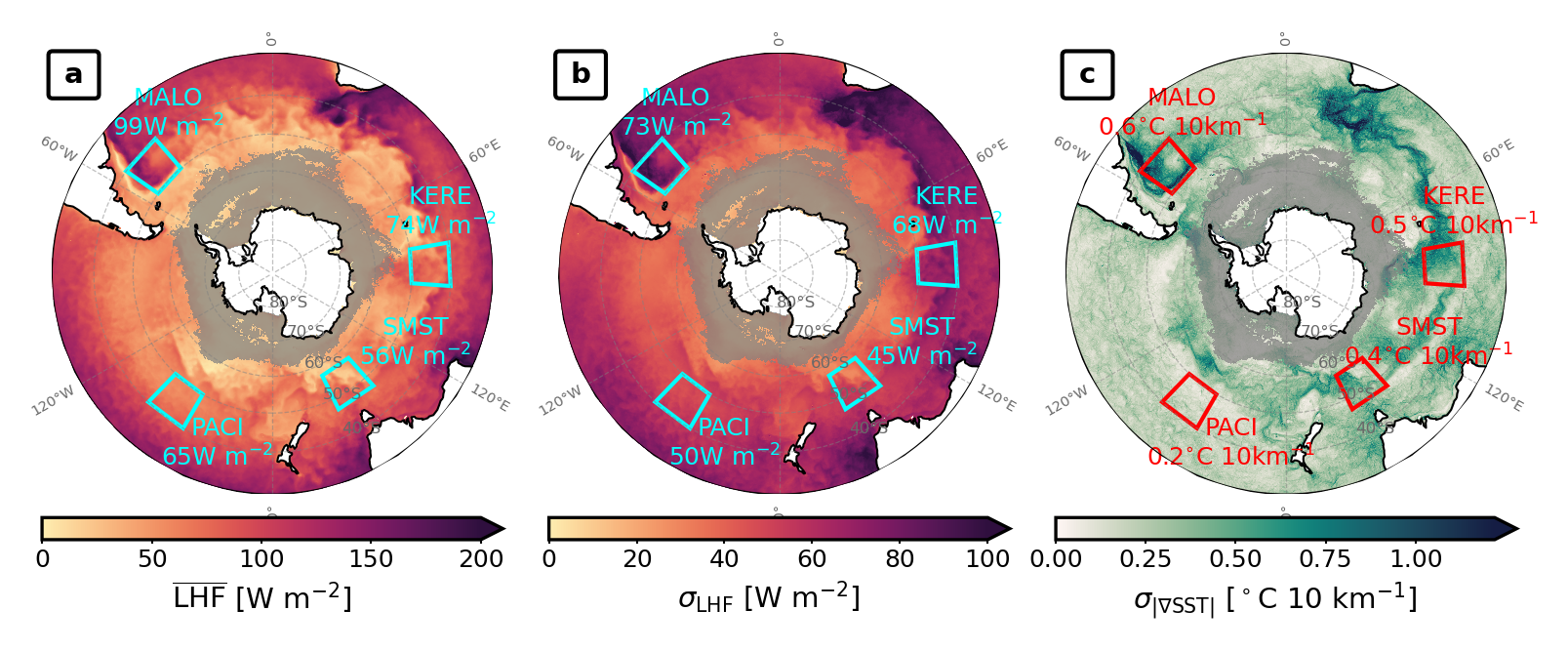}
 \caption{\textbf{Additional regions for spectral gap fitting in COAS.} Same as Fig.~\ref{fig-SOlhf}, with boxes showing the four regions used in addition to AGUL and CHIL to fit the spectral gap (see also Fig.~\ref{fig-PSDlhf-otherregions} and section~\ref{method-spectral-gap}).}
 \label{fig-6regions}
\end{figure}

The final fit is derived from the annual-mean PSD averaged over six regions (AGUL and CHIL as in Fig.~\ref{fig-PSDlhf} and MALO, KERE, SMST and PACI as in appendix Fig.~\ref{fig-PSDlhf-otherregions}). These regions of approximately $\text{1100 km}\times\text{1100 km}$ in size are distributed across the Southern Ocean (Fig.~\ref{fig-SOlhf} and appendix Fig.~\ref{fig-6regions}) in locations where we observe different levels of LHF variability and which are also known for having different levels of ocean kinetic energy \citep{thompson2014equilibration}. Note that all six regions are ice-free in COAS and, while sea-ice covered regions might exhibit different spectra owing to different air-sea-ice interaction processes than in the open ocean, we do not investigate these effects in this work. These limitations are discussed in the conclusion.

An analysis of the sensitivity of the variance partition to the linear fit was performed, where the percentage of LHF variance contained below the partition line was evaluated using the fit coefficients obtained natively in each region and those obtained after averaging across the six regions. The difference is $\lesssim$2\% in the four most energetic regions (AGUL, MALO, KERE and SMST), and $\lesssim$7\% in the two regions (CHIL, PACI) of weakest variance (not shown), providing confidence in our spectral partitioning.

\begin{figure}
 \centering
 \noindent\includegraphics{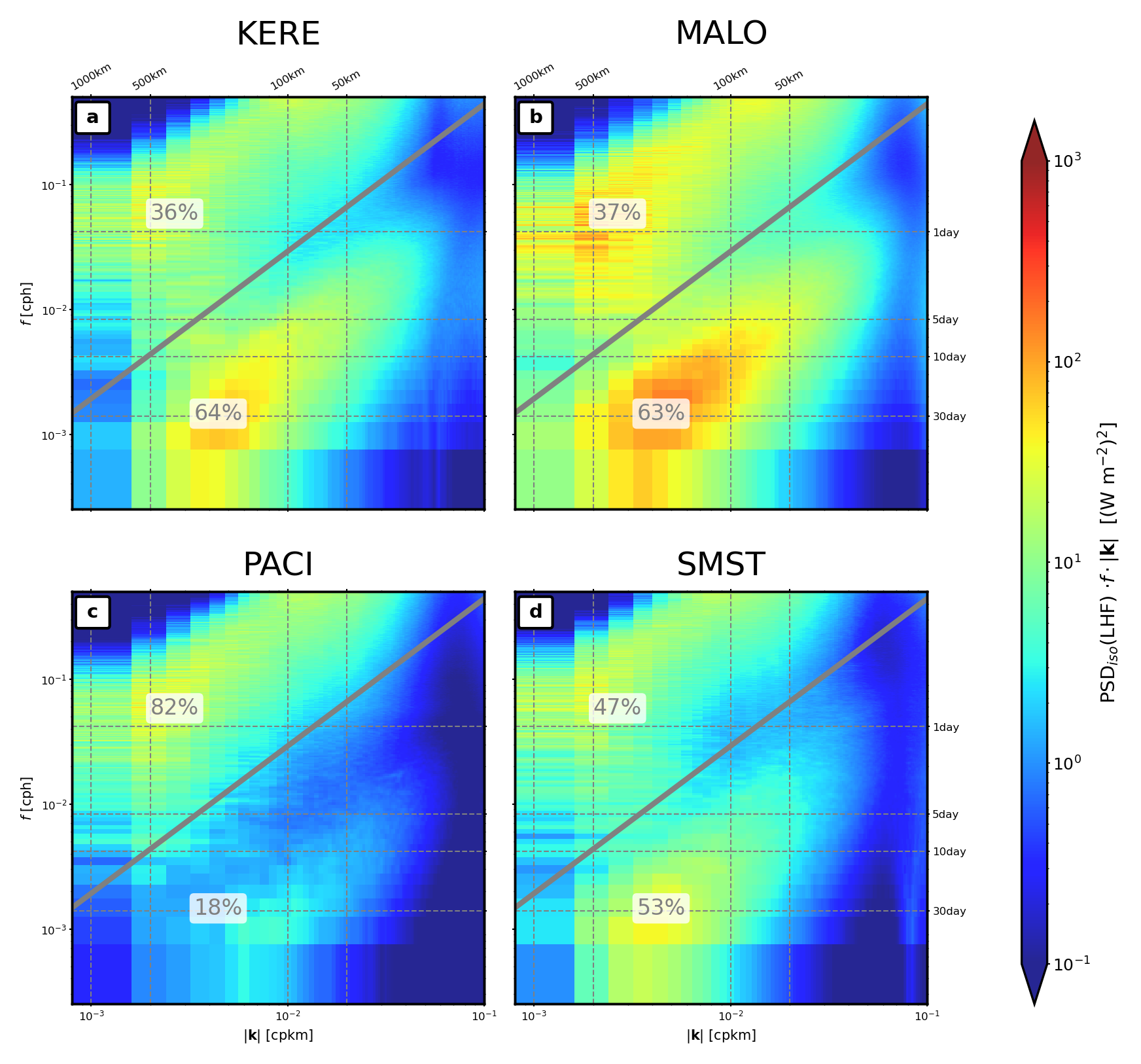}
 \caption{\textbf{Latent heat flux spectra in four additional regions in COAS.} Same as Fig.~\ref{fig-PSDlhf} for the four additional regions shown in Fig.~\ref{fig-6regions}: (a) KERE, (b) MALO, (c) PACI, and (d) SMST.}
 \label{fig-PSDlhf-otherregions}
\end{figure}

\clearpage

\appendix[C]

\appendixtitle{Joint distributions and conditional means}

Consider scalar fields $F(x,y,t), \alpha(x,y,t), \beta(x,y,t)$ defined in a domain $\mathscr{A}$. For clarity, we drop the time variable. Noting latitude as $\phi$, the mean of $F$ conditioned on $\alpha,\beta$ is defined as:
\begin{equation*}
\begin{split}
\bar{F}^{\alpha,\beta}
&=
\frac{\iint_A F(x,y)\,\delta[\alpha'(x,y)-\alpha]\,\delta[\beta'(x,y)-\beta]\,dx\,dy}{\iint_A \delta[\alpha'(x,y)-\alpha]\,\delta[\beta'(x,y)-\beta]\,dx\,dy}\\
&=
\frac{\sum_{(i,j)\in B_{\alpha\beta}} F_{ij}\,\Delta x\,\Delta y}{\sum_{(i,j)\in B_{\alpha\beta}} \Delta x\,\Delta y}\\
&=
\frac{\sum_{(i,j)\in B_{\alpha\beta}} F_{ij}\;\cos{\phi_{j}}}{\sum_{(i,j)\in B_{\alpha\beta}} \cos{\phi_{j}}}
\end{split}
\end{equation*}

\begin{align*}
B_{\alpha\beta} = \Bigl\{ (i,j) \; 
  & \alpha_{ij} \in [\alpha - d\alpha/2, \alpha + d\alpha/2), \\
  & \beta_{ij} \in [\beta - d\beta/2, \beta + d\beta/2) 
\Bigr\}
\end{align*}

The second line is a generic expression on a discrete, regularly spaced grid. In this study, the third expression is used where the grid cell area is proportional to the cosine of latitude $\phi$.

The joint probability of two scalar fields $\alpha,\beta$ is defined as:
\begin{equation*}
\begin{split}
p^{\alpha,\beta} 
&=
\frac{\iint_A \delta[\alpha'(x,y)-\alpha]\,\delta[\beta'(x,y)-\beta]\,dx\,dy}{\iint_A dx\,dy}\\
&=  \frac{\sum_{(i,j)\in B} \cos{\phi_j}}{\sum_{(i,j)) \in A} \cos{\phi_j}}
\end{split}
\end{equation*}

\bibliographystyle{ametsocV6}
\bibliography{references}

\end{document}